%% file: Paper.tex
\documentclass{sig-alternate}
\usepackage{latexsym}
\usepackage{times}
\usepackage{epsfig}
\usepackage{subfigure}
\usepackage{multirow}
\usepackage{amsfonts}
\usepackage{amsmath}
\usepackage{caption}
\usepackage{algorithm,algorithmic}
\usepackage{color}

\input{preamble.tex}

        {\medskip}

        {\hspace*{\fill}$\Box$\par\vspace{4mm}}
        {\hspace*{\fill}$\Box$\par}

\title{Limiting the Neighborhood: De-Small-World Network for Outbreak Prevention}

\numberofauthors{2}
\author{
\alignauthor
Ruoming Jin$^\dagger$~~~~~~~~~~~~~~~Yelong Sheng$^\dagger$\\
       \affaddr{$^\dagger$\mbox{ }Department of Computer Science}\\
       \affaddr{Kent State University}\\
       \email{\{jin,ysheng,lliu\}@cs.kent.edu}
\alignauthor
Lin Liu$^\dagger$~~~~~~~~~~~~~~~Xue-Wen Chen$^\ddagger$\\
       \affaddr{$^\ddagger$\mbox{ }Department of EECS}\\
       \affaddr{The University of Kansas}\\
       \email{xwchen@ku.edu}
}  

\begin{document}

\maketitle
\input{abs.tex}
\input{intro.tex}

\input{problem.tex}
\input{optimization.tex}

\input{speedup.tex}

\input{exp.tex}

\input{conclude.tex}

\begin{small}
\bibliographystyle{abbrv}
\bibliography{bib/paper}
\end{small}
\input{Appendix.tex}

\end{document}

%% file: preamble.tex
\newcommand{\comment}[1]{}

\makeatletter
\newcommand{\rmnum}[1]{\romannumeral #1}
\newcommand{\Rmnum}[1]{\expandafter\@slowromancap\romannumeral #1@}
\makeatother

\newsavebox{\boxone}
\newsavebox{\boxtwo}
\newsavebox{\boxthree}

\newlength{\narrow}
\newlength{\cnarrow}
\newcommand{\topline}{
  \hrule
  \vskip .5\baselineskip}
\newcommand{\bottomline}{
  \vskip 2pt
  \hrule}

\newcommand{\chbox}[2]{
  \hbox to #1{\hss\vtop{#2}\hss}}

\newcommand{\nchbox}[1]{
  \chbox{\narrow}{#1}}

\newcommand{\cnchbox}[1]{
  \chbox{\cnarrow}{#1}}

\newcommand{\fcode}[1]{
  
  \chbox{\textwidth}{\tgrind\input{#1}}}

\newcommand{\ncode}[1]{
  
  \chbox{\narrow}{\tgrind\input{#1}}}

\def\nfig#1#2#3{
  \vtop{\nchbox{#1}
  \hbox to\narrow{\parbox{\narrow}{\caption{#2}\label{#3}}}}}

\newcommand{\cncode}[1]{
  \chbox{\cnarrow}{\tgrind\input{#1}}}

\def\codefiggen[#1]#2#3#4#5#6{
  \begin{figure}[#1]
  #5
  \fcode{#2}
  \center\parbox{.9\textwidth}{\caption{#3}\label{#4}}
  #6
  \end{figure}}

\def\codefig[#1]#2#3#4{
  \codefiggen[#1]{#2}{#3}{#4}{}{}}

\def\codefigline[#1]#2#3#4{
  \codefiggen[#1]{#2}{#3}{#4}{\topline}{\bottomline}}

\def\doublefiggen[#1]#2#3#4#5#6#7#8#9{
  \begin{figure}[#1]
  #8
  \hbox to \textwidth{
  \nfig{#2}{#3}{#4}
  \hfil
  \nfig{#5}{#6}{#7}}
  #9
  \end{figure}}

\def\doublefig[#1]#2#3#4#5#6#7{
  \doublefiggen[#1]{#2}{#3}{#4}{#5}{#6}{#7}{}{}}

\def\doublefigline[#1]#2#3#4#5#6#7{
  \doublefiggen[#1]{#2}{#3}{#4}{#5}{#6}{#7}{\topline}{\bottomline}}

\def\doublecodefig[#1]#2#3#4#5#6#7{
  \doublefig[#1]{\ncode{#2}}{#3}{#4}{\ncode{#5}}{#6}{#7}}

\def\doublecodefigline[#1]#2#3#4#5#6#7{
  \doublefigline[#1]{\ncode{#2}}{#3}{#4}{\ncode{#5}}{#6}{#7}}

\newcommand{\codepair}[4]{\vbox{
  \hbox{\ncode{#1} \hfil \ncode{#3}}
  \vskip .3\baselineskip plus .3\baselineskip
  \hbox{\hbox to\narrow{#2\hfil} \hfil \hbox to\narrow{#4\hfil}}}}

\def\codepairfig[#1]#2#3#4#5#6#7{
  \begin{figure}[#1]
  \codepair{#2}{#3}{#4}{#5}
  \center\parbox{.9\textwidth}{\caption{#6}}
  \label{#7}
  \end{figure}}

\def\cncodepairfiggen[#1]#2#3#4#5#6#7{
  \begin{figure}[#1]
  #6
  \hbox{\cncode{#2}\hfil\cncode{#3}}
  \center\parbox{.9\columnwidth}{\caption{#4}\label{#5}}
  #7
  \end{figure}}

\def\cncodepairfig[#1]#2#3#4#5{
  \cncodepairfiggen[#1]{#2}{#3}{#4}{#5}{}{}}

\def\cncodepairfigline[#1]#2#3#4#5{
  \cncodepairfiggen[#1]{#2}{#3}{#4}{#5}{\topline}{\bottomline}}

\def\doublefigOnecap*[#1]#2#3#4#5{
  \begin{figure*}[#1]
  \hbox to \textwidth{
  \nchbox{#2}
  \hfil
  \nchbox{#3}}
  \caption{#4}
  \label{#5}
  \end{figure*}}

\def\doublefigOnecap[#1]#2#3#4#5{
  \begin{figure}[#1]
  \topline
  \hbox to \columnwidth{
  \cnchbox{#2}
  \hfil
  \cnchbox{#3}}
  \caption{#4}
  \label{#5}
  \bottomline
  \end{figure}}

\def\PSfig[#1]#2#3#4{
 \begin{figure}
 \centerline{\psfig{file=#2,width=\columnwidth}}
 \caption{{#3}}
 \label{#4}
 \end{figure}}

\def\PSfiglines[#1]#2#3#4{
 \begin{figure}[#1]
 \topline
 \centerline{\psfig{file=#2,width=\columnwidth}}
 \caption{{#3}}
 \label{#4}
 \bottomline
 \end{figure}}

\def\PSfiglinesht[#1]#2#3#4#5{
 \begin{figure}[#1]
 \topline
 \centerline{\psfig{file=#2,height=#3}}
 \caption{{#4}}
 \label{#5}
 \bottomline
 \end{figure}}

\def\doublePSfig[#1]#2#3#4#5#6{
  \doublefigOnecap[#1]
    {\cnchbox{\psfig{file=#2,height=#4}}}
    {\cnchbox{\psfig{file=#3,height=#4}}}
    {#5}
    {#6}}

\newlength{\boxwidth}
\newcommand{\bproof}{{\bf Proof Sketch: }}
\newcommand{\eproof}{\mbox{$\Box$}}

\def\tabdoublecode#1#2#3#4{
 \begin{figure*}[t]
 \topline\vs{-.4}
 \hbox to \columnwidth{
 \vtop{\small
 \begin{tabbing}
 #1
 \end{tabbing}}
 \hfil
 \hfil
 \hfil
 \vtop{\small
 \begin{tabbing}
 #2
 \end{tabbing}}
 }
 \caption{#3\label{#4}}
 \bottomline
 \end{figure*}
}
\def\tabtriplecode#1#2#3#4#5{
 \begin{figure}
 \topline\vs{-.4}
 \hbox to \columnwidth{
 \vtop{\small
 \begin{tabbing}
 #1
 \end{tabbing}}
 \hfil
 \hfil
 \hfil
 \vtop{\small
 \begin{tabbing}
 #2
 \end{tabbing}}
 \hfil
 \hfil
 \hfil
 \vtop{\small
 \begin{tabbing}
 #3
 \end{tabbing}}
 }
 \caption{#4\label{#5}}
 \bottomline
 \end{figure}
}

\newtheorem{lemma}{Lemma}
\newcommand{\blemma}{\begin{lemma}}
\newcommand{\elemma}{\end{lemma}}

\newtheorem{thm}{Theorem}
\newcommand{\bthm}{\begin{thm}}
\newcommand{\ethm}{\end{thm}}

\newtheorem{defin}{Definition}
\newcommand{\bdefin}{\begin{defin}}
\newcommand{\edefin}{\end{defin}}

\newtheorem{observation}{Observation}
\newcommand{\bobserv}{\begin{observation}}
\newcommand{\eobserv}{\end{observation}}

\newcommand{\vs}[1]{\vspace{#1cm}}
\newcommand{\be}{\begin{equation}}
\newcommand{\ee}{\end{equation}}

\newcommand{\bdesc}{\begin{description}}
\newcommand{\edesc}{\end{description}}
\newcommand{\benum}{\begin{enumerate}}
\newcommand{\eenum}{\end{enumerate}}
\newcommand{\bitem}{\begin{itemize}}
\newcommand{\eitem}{\end{itemize}}
\newcommand{\bcenter}{\begin{center}}
\newcommand{\ecenter}{\end{center}}
\newcommand{\btabular}{\begin{tabular}}
\newcommand{\etabular}{\end{tabular}}
\newcommand{\beqnarr}{
 \begin{eqnarray}}
\newcommand{\eeqnarr}{\end{eqnarray}}

%% file: abs.tex
\begin{abstract}
In this work, we study a basic and practically important strategy to help
prevent and/or delay an outbreak in the context of network: limiting the
contact between individuals. In this paper, we introduce the 
{\it average neighborhood size} as a new measure for the degree of 
being small-world and utilize it to formally define the {\em de-small-world} network
problem. We also prove the NP-hardness of the general reachable pair cut problem
and propose a greedy edge betweenness based approach as the benchmark in selecting
the candidate edges for solving our problem. 
Furthermore,  we transform the de-small-world network problem as an 
OR-AND Boolean function maximization problem, which is 
also an NP-hardness problem. In addition, we develop a numerical relaxation approach to
solve the Boolean function maximization and the de-small-world problem.  Also, we
introduce the {\it short-betweenness}, which measures the edge importance in terms
of all short paths with distance no greater than a certain threshold, and utilize it to speed up
our numerical relaxation approach. The experimental evaluation demonstrates the effectiveness and efficiency 
of our approaches.
\end{abstract}

%% file: intro.tex
\section{Introduction}

The interconnected network structure has been recognized to play a pivotal role in many complex systems, ranging from natural (cellular system), to man-made (Internet), to the social and economical systems.  
Many of these networks exhibit the ``small-world'' phenomenon, i.e., any two vertices in the network is often connected by a small number of intermediate vertices (the shortest-path distance is small). The small-world phenomenon in the real populations was first discovered by Milgram~\cite{Milgram67}.  In his study, the average distance between two Americans is around $6$. Several recent studies ~\cite{Leskovec:2008:PVL, Mislove:2007:MAO,Jin:2006:SCI} offer significant evidence to support similar observations in the online social networks and Internet itself. In addition, the power-law degree distribution (or scale-free property) which many of these networks also directly lead to the small average distance~\cite{Andersen_Chung_Lu_2005}.
Clearly, the small-world property can help facilitate the communication and speed up the diffusion process and information spreading in a large network.

However, the small-world effect can be a dangerous {\em double-edged sword}. 
When a system is benefited from the efficient communication and fast information diffusion, it also makes itself more vulnerable to various attacks: diseases, (computer) virus, spams, and misinformation, etc. 
For instance, it has been shown that a small-world graph can have much faster disease propagation than a regular lattice or a random graph~\cite{citeulike:5880645}. Indeed, the six degrees of separation may suggest that a highly infectious disease could spread to all six billion people living in the earth about size incubation periods of the diseases~\cite{citeulike:5880645}. 
The small-word property of Internet and WWW not only enables the computer virus and spams to be much easier to spread, but also makes them hard to stop. 
More recently, the misinformation problem in the social networks has made several public outcry~\cite{Budak:2011:LSM}.
These small-world online  social network potentially facilitate the spread of misinformation to reach a large number of audience in short time, which may cause public panic and have other disruptive effects. 

To prevent an outbreak, the most basic strategy is to remove the affected individuals (or computers) from the network system, like quarantine. 
However, in many situations, the explicit quarantine may be hard to achieve: the contagious individuals are either unknown or hard to detect; or it is often impossible to detect and remove each infected individual; or there are many already being affected and it become too costly to remove all of them in a timely fashion. 
Thus, it is important to consider alternative strategies to help prevent and even delay the spreading where the latter can be essential in discovering and/or deploying new methods for dealing with the outbreaks. 

Recently, there have been a lot of interests in understanding the network factors (such as the small-word and scale-free properties) in the epidemics and information diffusion process, and utilizing the network structures in detecting/preventing the outbreaks. 
Several studies have focused on modeling the disease epidemics on  the small-world and/or scale-free networks \cite{PhysRevLett.86.3200, citeulike:5880645}, \cite{citeulike:3053277};  in ~\cite{Leskovec:2007:COD}, Leskovec {\em et al.} study how to deploy sensors cost-effectively in a network system (sensors are assigned to vertices) to detect an outbreak; in ~\cite{Budak:2011:LSM}, Budak {\em et al.} consider how to limit the misinformation by identifying a set of individuals that are needed to adopt the ``good'' information (being immune in epidemics) in order to minimize those being affected by the ``bad'' information (being infected in epidemics). 
In addition, we note that from a different angle (viral marketing), there have been a list of studies on the {\em influence maximization} problem~\cite{Richardson:2002:MKS,Kempe:2003:MSI}, which aim to discover a set of most influential seeds to maximize the information spreading in the network. From the disease epidemics perspective, those seeds (assuming being selected using contagious model) may need particular protection to prevent an outbreak. 

In this work, we study another basic and practically important strategy to help prevent and/or delay an outbreak in the context of network: limiting the contact between individuals. 
Different from the pure quarantine approach, here individuals can still perform in the network system, though some contact relationships are forbidden. 
In other words, instead of removing vertices (individuals) form a network as in the quarantine approach, this strategy focuses on removing edges so that the (potential) outbreaks can be slowed down. 
Intuitively, if an individual contacts less number of other individuals, the chance for him or her to spread or being infected from the disease (misinformation) becomes less. 
From the network viewpoint, the edge-removal strategy essentially make the underlying (social) network less small-world, or simply ``de-small-world'', i.e., the distances between individuals increase to delay the spreading process. 
In many situations, such a strategy is often easily and even voluntarily adopted.
For instance, during the SARS epidemic in Beijing, 2004, there are much less people appearing in the public places. 
In addition, this approach can also be deployed in complement to the quarantine approach. 
 
\subsection{Our Contribution}
Even though the edge-removal or {\em de-small-world} approach seems to be conceptually easy to understand, its mathematical foundation is still lack of study. 
Clearly, different edges (interactions) in the network are not being equivalent in terms of slowing down any potential outbreak:  for a given individual, a link to an individual of high degree connection can be more dangerous than a link to another one with low degree connection. 
The edge importance (in terms of  distance) especially coincides with Kleighnberg's theoretical model~\cite{citeulike:3158434} which utilizes the {\em long-range edges} on top of an underlying grid for explaining the  small-world  phenomenon. 
In this model, the long-range edges are the main factors which help connect the otherwise long-distance pairs with a smaller number of edges.  
However, there are no direct studies in fitting such a model to the real world  graph to discover those long-range edges. 
In the mean time, additional constraint, such as the number of edges can be removed from the network, may exist because removing an edge can associate with certain cost. 
These factors and requirements give arise to the following fundamental research problem: {\em how can we maximally de-small-world a graph (making a graph to be less small-world) by removing a fixed number of edges?}

To tackle the {\em de-small-world} network problem, we make the following contributions in this work: 
\benum 
\item We introduce the {\em average neighborhood size} as a new measure for the degree of being small-world and utilizes it to formally define the de-small-world network problem. Note that the typical average distance for measuring the small-world effects cannot uniformly treat the connected and disconnected networks;  neither does it fit well  with the spreading process. We also reformulate the de-small-world as the {\em local-reachable pair cut} problem. 

\item We prove the NP-hardness of the general reachable pair cut problem and propose a greedy edge betweenness based approach as the benchmark in selecting the candidate edges  for  solving the de-small-world network. We transform  the de-small-world network problem and express it as a OR-AND Boolean function maximization problem, which is also an NP-hard problem. 

\item We develop a numerical relaxation approach to solve the de-small-world problem using its OR-AND boolean format. Our approach can find a local minimum based on the iterative gradient optimization procedure. In addition, we further generalize the betweenness measure and introduces the {\em short betweenness}, which measures the edge importance in terms of all the paths with distance no greater than a certain threshold. Using this measure, we can speed up the numerical relaxation approach by selecting a small set of candidate edges for removal. 

\item   We perform a detailed experimental evaluation, which demonstrates the effectiveness and efficiency of proposed approaches. 

\eenum 


\comment{
Furthermore, in order to model the {\em de-small-world} or edge-removal process, a criterion is needed to precisely capture the {\em degree} of small-world. 
The typical measure is based on the average distance. However, this measure is not able to unify 

 capture the effects of

Anderson {\em et al.} ~\cite{} propose a hybrid model which combines a global power-law graph with a local graph (with high local connectivity) for modeling both small-average distance and the clustering effect (two vertices with common neighborhoods tend to be adjacent).

This give arise to the following fundamental research problem: how can we transform a graph of being small-world to de-small-world? 
We also note that this problem is different from the clustering problem, which tries to remove certain edges so that the entire graph is decomposed into small components. 
Note that each small component can be still small-word. 
Thus, the general average distance may not be  a good criteria to characterize such a transformation. 
}

\comment{
the average of {em average distance} is small and the 
Network structure enables the communication, information 
When a large-scale infectious disease epidemic occurs, what kind of measures the government should take to reduce the number of infected peoples?
i.e. SARS epidemic in Beijing, 2004. Does social network analysis can tell us which is the most efficient way to control the disease spreading? }

%% file: problem.tex
\section{Problem Definition and Preliminary}
\label{problem}
In this section, we first formally define the {\em de-small-world} network problem and  prove its NP-hardness in (Subsection~\ref{definition}); then we introduce the basic greedy approaches based on edge betweenness which will serve as the basic benchmark (Subsection~\ref{benchmark}); and finally we show the de-small-world network problem can be transformed and expressed as a OR-AND Boolean function maximization problem (Subsection~\ref{ANDOR}). 


\subsection{Problem Formulation}
\label{definition}
In order to model the edge-removal process and formally define the {\em de-small-world} network problem, a criterion is needed to precisely capture the {\em degree} of being small-world. 
Note that here the goal is to help prevent and/or delay the potential outbreak and epidemic process. 
The typical measure of small-world network is based on the average distance (the average length of the shortest path between any pair of vertices in the entire network). 
However, this measure is not able to provide unified treatment of  the connected and cut network. 
Specifically, assuming a connected network is broken into several cut network and the average distance on the cut network is not easy to express. 
On the other hand, we note that the de-small-world network graph problem is different from the network decomposition (clustering) problem which tries to break the entire network into several components (connected subgraphs). 
From the outbreak prevention and delaying perspective, the cost of network decomposition is not only too high, but also may not be effective. 
This is because each individual component itself may still be small-world; and the likelihood of completely separating the contagious/infected group from the rest of populations (the other components) is often impossible. 

Given this, we introduce the {\em average neighborhood size} as a new measure for the degree of being small-world and utilize it to formally define the de-small-world network problem.   
Especially, the new measure can not only uniformly treat both connected and cut networks and aims to directly help model the spreading/diffusion process. 
Simply speaking, for each vertex $v$ in a network $G=(V,E)$ where $V$ is the vertex set and $E$ is the edge set, we define the neighborhood of $v$ as the number of vertices with distance no greater than $k$ to $v$, denoted as $N^{k}(v)$. 
Here $k$ is the user-specified {\em spreading} (or delaying) parameter which aims to measure the outbreak speed, i.e., in a specified time unit, the maximum distance between individual $u$ (source) to another one $v$ (destination) who can be infected if $u$ is infected. 
Thus, the average neighborhood size of $G$, $\sum_{v \in V}N^k(v)$, can be used to measure the robustness of the network with respect to a potential outbreak in a certain time framework. 
Clearly, a potential problem of the small-world network is that even for a small $k$, the average neighborhood size can be still rather large, indicating a large (expected) number of individuals can be quickly affected (within time framework $k$) during an outbreak process. 

Formally, the {\em de-small-world} network problem is defined as follows:
\bdefin({\bf De-Small-World Network Problem}) Given the edge-removal budget $L>0$ and the spreading parameter $k>1$ we seek a subset of edges $E_r \subset E$, where $|E_r| = L$, such that the average neighborhood size is minimized:  
\begin{eqnarray}
\min_{|E_r|=L} \frac{\sum_{v \in V} N^k(v|G\backslash E_r)}{|V|}, 
\end{eqnarray}
where $N^k(v|G\backslash E_r)$ is the neighborhood size of $v$ in the graph $G$ after removing all edges in $E_r$ from the edge set $E$. 
\label{def:problem_def}
\edefin 

Note that in the above definition, we assume each vertex has the equal probability to be the source of infection. 
In the general setting, we may consider to associate each vertex $v$ with a probability to indicate its likelihood to be (initially) infected. Furthermore, we may assign each edge with a weight to indicate the cost to removing such an edge. 
For simplicity, we do not study those extensions in this work; though our approaches can be in general extended to handle those additional parameters. 
In addition, we note that in our problem, we require the spreading parameter $k>1$. 
This is because for $k=1$, this problem is trivial: the average neighborhood size is equivalent to the average vertex degree; and removing any edge has the same effect. 
In other words, when $k=1$, the neighborhood criterion does not capture the spreading or cascading effects of the small-world graph. 
Therefore, we focus on $k>1$, though in general $k$ is still relatively small (for instance, no higher than $3$ or $4$ in general). 

\noindent{\bf Reachable Pair Cut Formulation:} 
We note the de-small-world network problem can be  defined in terms of the {\em  reachable pair cut} formulation. 
Let a pair of two vertices whose distance is no greater than $k$ is referred to as a {\it local-reachable pair} or simply {\em reachable pair}. Let $\mathcal{R}_G$ record the set of all local reachable pairs in $G$.  

\bdefin({\bf Reachable Pair Cut Problem})
For a given local $(u,v)$, if $d(u,v|G)\leq k$ in $G$, but $d(u,v|G\backslash E_s) >k$, where $E_s$ is an edge set in $G$, then we say $(u,v)$ is being {\bf local cut} (or simply cut) by $E_s$. 
Given the edge-removal budget $L>0$ and the spreading parameter $k>1$, the reachable pair cut problem aims to find the edge set $E_r \subseteq E$, such that the maximum number of pairs in $\mathcal{R}_G$ is cut by $E_r$. 
\edefin

Note that here the (local) cut for a pair of vertices simply refers to increase their distance; not necessarily completely disconnect them in the graph ($G \backslash E_s$).
Also, since $\mathcal{R}_{G \setminus E_r} \subseteq \mathcal{R}_{G}$, i.e., every local-reachable pair in the remaining network $G \setminus E_r$ is also the local-reachable in the original graph $G$, the problem is equivalently to maximize 
$|\mathcal{R}_{G}|- |\mathcal{R}_{G \setminus E_r}|$ and minimize the number of local reachable pairs $|\mathcal{R}_{G \setminus E_r}|$.  Finally, the correctness of such a reformulation  (de-small-world problem=reachable pair cut problem)  follows this simple observation: $\sum_{v \in V} N^k(v|G) = 2|\mathcal{R}_{G}|$ (and  $\sum_{v \in V} N^k(v|G \backslash E_r) = 2|\mathcal{R}_{G \setminus E_r}|$). 
Basically, every reachable pair is counted twice in the neighborhood size criterion. 

In the following, we study the hardness of the general reachable pair cut  problem.  
\bthm
\label{thm:nphard}
Given a set $RS$ of local reachable pairs in $G=(V, E)$ with respect to $k$, the problem of finding $L$ edges $E_r\subseteq E$ ($|E_r|=L$) in $G$ such that the maximal number of pairs in $RS$ being cut by $E_r$ is NP-Hard.  
\ethm

Note that in the general problem, $RS$ can be any subset of $\mathcal{R}_G$. 
The NP-hardness of the general reachable pair cut problem a strong indicator that the de-small-world network problem is also hard. The proof of Theorem~\ref{thm:nphard} is in Appendix. 
In addition, we note that the submodularity property plays an important role in solving vertex-centered maximal influence~\cite{Kempe:2003:MSI}, outbreak detection ~\cite{Leskovec:2007:COD}, and limiting misinformation spreading ~\cite{Budak:2011:LSM} problems. 
However, such property does not hold for the edge-centered de-small-world problem. 

\begin{figure} [!htp]
\vspace{-1ex}
\centering
\includegraphics[height=0.7in,width=1.5in]{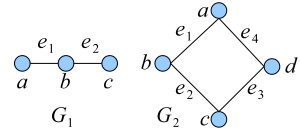}  
\caption{Counter Examples} 
\label{fig:counter_example}
\vspace{-3ex} 
\end{figure}

\blemma
\label{lemma:subsup}
Let set function $f: 2^E \rightarrow Z^+$ records the number of local reachable pairs in $\mathcal{R}_G$ is cut by an edge set $E_s$ in graph $G$. Function $f$ is neither submodular (diminishing return) nor supermodular. 
\elemma
\bproof
For a supermodular function: we have $f(A \cup B)+ f(A \cap B) \geq f(A) + f(B)$; 
for a submodular function: we need show $f(A)+f(B) \geq f(A \cup B) + f(A \cap B)$. 
Here we use two counter examples (suppose $k=2$). 
For $G_1$ in Figure \ref{fig:counter_example}, we can see that edge sets $\{e_1\}$ and $\{e_2\}$ cut (cut) respectively two reachable pairs $\{(ab, ac\}$ and $\{ac, bc\}$. Then we have $f(\{e_1\})+f(\{e_2\}) = 4$; however,
$f(\{e_1,e_2\})+f(\emptyset) = 3$. Therefore, supermodularity does not hold.
For $G_2$ in Figure \ref{fig:counter_example}, we can see that edge set $\{e_1\}$ and $\{e_3\}$
each can only cut one reachable pair. However, $\{e_1, e_3\}$ could cut four pairs. That means,
$f(\{e_1,e_3\})+f(\emptyset) > f(\{e_1\})+|f(\{e_3\})$. Therefore, submodularity can not hold.
\eproof 

\subsection{Greedy Betweenness-Based Approach}
\label{benchmark}
Finding the optimal solution for the de-small-world problem is likely to be NP-hard. 
Clearly, it is computationally prohibitive to enumerate all the possible removal edge set $E_r$ and to measure how many reachable pairs could be cut or how much the average neighborhood size is reduced. 
In the following, we describe a greedy approach to heuristically discover a solution edge-set.  This approach also serves as the benchmark for the de-small-world problem. 

The basic approach is based on the edge-betweenness, which is a useful criterion to measure the edge importance in a network. 
Intuitively, the edge-betweenness measures the edge important with respect to the shortest paths in the network. 
The high betweenness suggests that the edge is involved into many shortest paths; and thus removing them will likely increase the distance of those pairs linked by these shortest paths. 
Here, we consider two variants of edge-betweenness: the (global) edge-betweenness~\cite{GirvanNewman02} and the local edge-betweenness~\cite{Gregory08}. 
The global edge-betweenness is the original one~\cite{GirvanNewman02} and is defined as follows: 
\begin{eqnarray*}
B(e) = \sum_{s \neq t \in V} \frac{\delta_{st}(e)}{\delta_{st}},
\end{eqnarray*} 
where $\delta_{st}$ is the total number of shortest paths between vertex $s$ and $t$, and $\delta_{st}(e)$ the 
total number of shortest paths between $u$ and $v$ containing edge $e$.

The local edge-betweenness considers only those vertex pairs whose shortest paths are no greater than $k$, and is defined as 
\begin{eqnarray*}
LB(e) = \sum_{s \neq t \in V, d(s, t) \leq k} \frac{\delta_{st}(e)}{\delta_{st}},
\end{eqnarray*} 
The reason to use the local edge-betweenness measure is because in the de-small-world (and reachable pair cut) problem, we focus on those local reachable pairs (distance no greater than $k$). 
Thus, the contribution to the (global) betweenness from those pairs with distance greater than $k$ can be omitted. 
The exact edge-betweenness can be computed in $O(nm)$ worst case time complexity~\cite{Brandes01afaster} where $n=|V|$ (the number of vertices) and $m=|E$ (the number of edges) in a given graph, though in practical the local one can be computed much faster.

Using the edge-betweenness measure, we may consider the following {\em generic procedure} to select the $L$ edges for $E_r$: 

\noindent{1)} Select the top $r < L$ edges into $E_r$, and remove those edges from the input graph $G$; 

\noindent{2)} Recompute the betweenness for all remaining edges in the updated graph $G$; 

\noindent{3)} Repeat the above procedure $\lceil L/r \rceil$ times until all $L$ edges are selected. 

Note that the special case $r=1$, where we select each edge in each iteration, the procedure is very similar to the Girvan-Newman algorithm~\cite{GirvanNewman02} in which they utilize the  edge-betweenness  for community discovery. Gregory ~\cite{Gregory08} generalizes it to use the local-edge betweenness. 
Here, we only consider to pickup $L$ edges and allow users to select the frequency to recompute the edge-betweenness (mainly for efficiency consideration). 
The overall time complexity of the betweenness based approach is $O(\lceil L/r \rceil nm)$ (assuming the exact betweenness computation is adopted).

\subsection{OR-AND Boolean Function and its Maximization Problem} 
\label{ANDOR}

In the following, we transform  the de-small-world network problem and express it as a OR-AND Boolean function maximization problem, which forms the basis for our optimization problem in next section. 
First, we will utilize the OR-AND graph to help represent the de-small-world (reachable pair cut) problem. 
Let us denote $P$ the set of all the short paths in $G$ that have length at most $k$. 

\begin{figure} [!htp] 
\vspace{-1.5ex}
\begin{center} 
\centering 
\subfigure[\small Example Graph]{   
\includegraphics[height=1 in,width=1.2in]{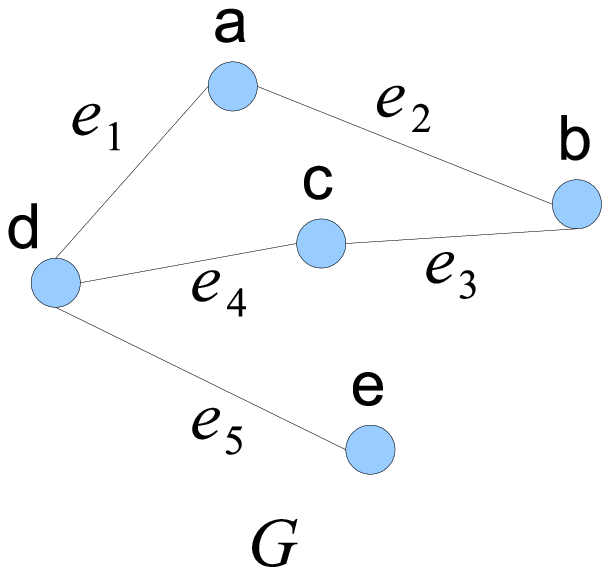}   
\label{fig:runningexample}  
} 
\subfigure[\small OR-AND Graph]{   
\includegraphics[height=1.5 in,width=1.5in]{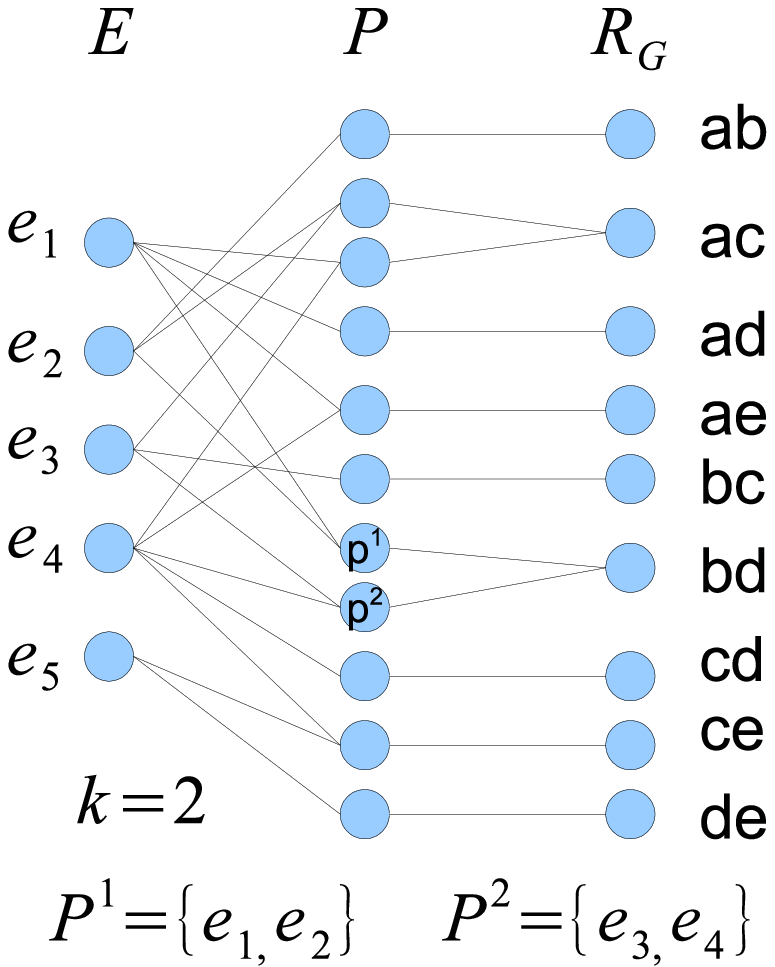}   
\label{fig:and_or}  
} 
\caption{{\it OR-AND} graph } 
\label{fig:runningexampleandor}
\end{center} 
\vspace{-3ex}  
\end{figure}   

\noindent{\bf OR-AND Graph:} 
Given a graph $G=(V,E)$, the vertex set of an {\it OR-AND} graph $\mathcal{G}=(\mathcal{V},\mathcal{E})$ 
is comprised of three kinds of nodes $\mathcal{V}_E$, $\mathcal{V}_P$ and $\mathcal{V}_{\mathcal{R}_G}$, where each node in $\mathcal{V}_E$ corresponds to a unique edge in $E$, each node in  $\mathcal{V}_P$ corresponds to a short path in $P$, and each node in $\mathcal{V}_{\mathcal{R}_G}$ corresponds to a unique reachable pair in $G$ (with respect to $k$). 
Figure~\ref{fig:and_or} shows those nodes for graph $G$ in Figure~\ref{fig:runningexample}. 
The edge set consists of two types of edges: 1) 
Each short path node in $\mathcal{V}_P$ is linked with the vertices in $\mathcal{V}_E$  corresponding to those edges in in the path.  For instance path node $p^1$ in $\mathcal{V}_P$ links to edge node $e_1$ and $e_2$ in  $\mathcal{V}_E$ in Figure~\ref{fig:and_or}. 
Each reachable pair node in $\mathcal{V}_{\mathcal{R}_G}$ links to those path nodes which connects the reachable pair . 
For instance, the reachable pair $bd$ is connected with path node $p^1$ and $p^2$ in Figure~\ref{fig:and_or}.

Intuitively, in the OR-AND graph, we can see that in order to cut a reachable pair, we have to cut {\it all} the short 
paths between them (AND).   To cut one short path, we need to remove only one edge in that path (OR). 
Let $P(u,v)$ consists all the (simple) short paths between $u$ and $v$ whose length are no more than $k$.
For each short path $p$ in $P(u,v)$, let $e$ corresponds to a Boolean  variable for edge $e \in p$: if $e_i=T$, then the edge $e_i$ is not cut; if $e_i=F$, then the edge is cut ($e_i \in E_r$). 
Thus, for each reachable pair $(u,v) \in \mathcal{R}_G$, we can utilize the a Boolean OR-AND expression to describe it: 
\beqnarr 
I(u,v)=\bigvee_{p \in P(u,v)} \bigwedge_{e \in p} e
\eeqnarr
For instance, in the graph $G$ (Figure~\ref{fig:and_or}), 
\beqnarr
I(b,d)=(e_1 \wedge e_2) \vee (e_3 \wedge e_4)  \nonumber
\eeqnarr
Here, $I(b,d)=T$ indicating the pair is being cut only if for both $p^1$ and $p^2$ are cut. 
For instance, if $e_1=F$ and $e_3=F$, then $I(b,d)=F$; and $e_1=F$, but $e_3=T$ and $e_4=T$, $I(b,d)=T$. 
Given this, the de-small-world problem (and the reachable pair cut problem) can be expressed as the following Boolean function maximization problem. 

\bdefin({\bf Boolean Function Maximization Problem})
{\em Given a list of Boolean functions (such as $I(u,v)$, where $(u,v) \in \mathcal{R}_G$), we seek a Boolean variable assignment where exactly $L$ variables are assigned false ($e=F$ iff $e \in E_r$, and $|E_r|=L$), such that the maximal number of Boolean functions being false ($I(u,v)=F$ corresponding to $(u,v)$ is cut by $E_r$).}
\edefin
Unfortunately, the Boolean function maximization problem is also NP-hard since it can directly express the general reachable pair cut problem. 
In the next section, we will introduce a numerical relation approach to solve this problem.

\comment{
Therefore, edges incident on the same node in $\mathcal{R}_G$ column stands for an {\it AND} relationship  
(because all short paths between the two vertices have to be cut spontaneously), and  
edges incident on the same node in $P$ column from $E$ column stands for a {\it OR} relationship(because 
for each path any its edges can cut it). That is the reason we call it {\it OR-AND} graph. 

For {\it OR-AND} graph, we have the following lemma. 

\blemma{\bf {\it OR-AND} Graph Property:}  
Given function$f:2^{\mathcal{V}} \rightarrow N$,   
given $\mathcal{V} = P$ or $\mathcal{V} = E$, $|f|$ is neither a submodular nor supermodular function. 
\label{lemma:and_or} 
\elemma 

For Lemma \ref{lemma:and_or}, counters examples can be easily given for submodularity or supermodularity.  
Compared with Lemma \ref{lemma:or_and}, we can not find submodularity for {\it OR-AND} graph. However, 
the benefit we get from short paths is that the number of short paths between two vertices is normally very 
small. For example, in Figure \ref{fig:and_or}, 
with $k = 2$, at most two short paths between any reachable pairs for $G$. 
Therefore, in this paper, it is reasonable for us to tackle our problem from the perspective of short paths.  

Note that {\it OR-AND} graph closely relates to both our algebriacal problem definition and short betweenness  
definition which will be discussed later in this section. 

}

\comment{
\subsection{Cut-Based Approach}

Even of their simplicity, the two greedy heuristics proposed in last subsection can not provide any 
gaurantee on the accuracy of their results. Therefore, from this subsection, we plan to theorectically 
study some properties of our problem, and develop efficient algorithms. 

From the problem definition, we find that our problem can be represented in two different graphs: 
{\it OR-AND} and {\it OR-AND} graphs. 
Conceptually, for graph $G$ in Figure \ref{fig:running_example}, its 
{\it OR-AND} graph and {\it OR-AND} graph are respectively in Figures \ref{fig:or_and} and \ref{fig:and_or}.
In this section, we will focus on {\it OR-AND} graph, and {\it OR-AND} graph will be discussed 
in next section.


Our goal is to reduce the number of reachable pairs by removing
$l$ edges. In other words, we intuitively want to remove the edges that form cuts for reachable pairs.
Our first intuition then is to find the edge set $E_r$ with candinality $l$ such that the cut sets
$\mathcal{C}$ formed by $E_r$ disconnects the maximum number of reachable pairs. 
{\it Here, a cut $C(u,v) \subset E$ means a length-bounded s-t cut for vertice $u$ and $v$; that is, 
$u$ and $v$ are not connected within length $k$ in graph $G\setminus C(u,v)$}. Hereafter, we 
use cut and length-bounded s-t cut interchangeably. To understand the relationship between cuts and edges,
we define {\it OR-AND} (or cut based) graph.

\bdefin{\bf OR-AND Graph:} OR-AND graph has three columns ($E, C, \mathcal{R}_G$) of nodes, respectively
standing for edge set, cut set and the set of reachable pairs. 
In $E$ column, each node represents one edge $e \in E(G)$; in $C$
column, each node stands for an length-bounded s-t edge cut. Each cut is connected
with the edges in $E$ column which consists of it. Meanwhile each cut also connected
to certain nodes in $\mathcal{R}_G$ column, which stands for reachable pairs removed by this cut.
Figure \ref{fig:or_and} gives us an example. 
\edefin 

For example, from Figure \ref{fig:or_and}, we can see that vertices $b$ and $d$ could be separated by
four different cuts: $C^1, C^2, C^3, C^4$, because the reachable pair $bd$ is connected
to four nodes in $C$ column, and that each cut $C^i$ consists of two edges, e.g. 
$C^1=\{e_2, e_3\}$, because the their corresponding nodes are respectively connected to two nodes 
in $E$ column.

Given {\it OR-AND} graph, what properties does our problem have? Let $\mathcal{V}$ be a finite set, 
and an interger-valued function $f:2^{\mathcal{V}} \rightarrow N$ is submodular if and 
only if it satisfies the submodular inequality,
\[
	f(A \cup B) + f(A \cap B) \leq f(A) + f(B),
\] 
for all subsets $A, B \subseteq V$.

For our problem, first if $\mathcal{V} = \mathcal{C} = \{C^1, C^2, \cdots, C^n\}$ is the set of 
cuts, 
and for $A\subseteq \mathcal{V}$, $f(A)$ stands for the set of reachable pairs that can be removed by $A$.
Then we can easily see that $|f|$ is a submodular function. However, instead if $\mathcal{V}=E$ and 
for any edge subset $A \subseteq \mathcal{V}$, $f(A)$ also stands for the set of reachable pairs removed 
by $A$, then $|f|$ is neither sub or super modular function. More formally, we have 

\blemma{\bf {\it OR-AND} Graph Property:} 
Given function$f:2^{\mathcal{V}} \rightarrow N$,  
1) given $\mathcal{V} = \{C^1, C^2, \cdots, C^n\}$, $|f|$ is a submodular function;
2) given $\mathcal{V} = E$, $|f|$ is neither a submodular nor supermodular function.
\label{lemma:or_and}
\elemma
\bproof
1) Given $\mathcal{V} = \{C^1, C^2, \cdots, C^n\}$, any $A, B \subseteq V$, 
when $A \cap B = \emptyset$,
it is easy to see that $|f(A \cup B)| \leq |f(A)| + |f(B)|$. Then for $A \cap B \neq \emptyset$, we
have two cases.
(\Rmnum{1}): $f(A\setminus B) \cap f(B \setminus A) = \emptyset$. Then there are four small cases. 
(a)$f(A\cap B) \cap f(A\setminus B) \neq \emptyset$ and  $f(A\cap B) \cap f(B\setminus A) = \emptyset$.
{\small
\begin{eqnarray*}
&&|f(A \cup B)| + |f(A \cap B)| \\
&=& |f(A\setminus B)\setminus f(A\cap B)|\\
&+& |f(B\setminus A)| + 2|f(A \cap B)|\\
&=& |f(A)| + |f(B)|.
\end{eqnarray*}
}
Similarly, we can prove that $|f(A \cup B)| + |f(A \cap B)| = |f(A)| + |f(B)|$ holds when 
(b) $f(A\cap B) \cap f(A\setminus B) = \emptyset$ and  $f(A\cap B) \cap f(B\setminus A) \neq \emptyset$, 
(c) $f(A\cap B) \cap f(A\setminus B) \neq \emptyset$ and  $f(A\cap B) \cap f(B\setminus A) \neq \emptyset$,
and (d) $f(A\cap B) \cap f(A\setminus B) = \emptyset$ and  $f(A\cap B) \cap f(B\setminus A) = \emptyset$.
Therefore, we can see that the relationship among $f(A\cap B)$, $f(B\setminus A)$ and $f(A \setminus B)$
will not effect the inequality. 
(\Rmnum{2}): $f(A\setminus B) \cap f(B \setminus B) \neq \emptyset$. Given the result of (\Rmnum{1}),
here, w.o.l.g., we only consider the case 
$f(A\setminus B) \cap f(A\cap B) = \emptyset$ and $f(A\setminus B) \cap f(A\cap B) = \emptyset$.
In this case, clearly we have $|f(A \cup B)| + |f(A \cap B)| < |f(A)| + |f(B)|$.

Combining (\Rmnum{1}) and (\Rmnum{2}), the proof for 1) is complete. 

2) Here we use two counter examples (suppose $k=2$). 
For $G_1$ in Figure \ref{fig:counterexample}, we can see that edge sets $\{e_1\}$ and $\{e_2\}$ cut respectively two reachable
pairs $\{(ab, ac\}$ and $\{ac, bc\}$. Then we have $|f(\{e_1\})|+|f(\{e_2\})| = 4$; however,
$|f(\{e_1,e_2\})|+|f(\emptyset)| = 3$. Therefore, supermodularity does not hold.
For $G_2$ in Figure \ref{fig:counterexample}, we can see that edge set $\{e_1\}$ and $\{e_3\}$
can only cut one reachable pair separately. However, $\{e_1, e_3\}$ could cut four pairs. That means,
$|f(\{e_1,e_3\})|+|f(\emptyset)| > |f(\{e_1\})|+|f(\{e_3\})|$. Therefore, submodularity can not hold.
\eproof


Given Lemma \ref{lemma:or_and}, in cut level, we could then easily develop efficient greedy algorithm
as follows.

\begin{algorithm}[ht]
\caption{${\bf GreedyCover}(\mathcal{R}_{G}, \mathcal{C})$} 
\begin{algorithmic}[1]
\label{alg:greedcover}
\STATE {\bf Input}: $\mathcal{R}_{G}$: the set of reachable pairs; 
\STATE {\bf Input}: $\mathcal{C}$: the set of cuts;
\STATE {\bf Output}: cut set $C_r$ removed;
\WHILE{ $|C_r| < k$ and $|\mathcal{R}_G| > 0$  }
\STATE $C = \arg_{C^i \in \mathcal{C}} \max f(\{C^i\})$;
\STATE $\mathcal{R}_G = \mathcal{R}_G \setminus f(\{C\})$;
\STATE $C_r = C_r \cup \{C\}$;
\ENDWHILE
\end{algorithmic}
\end{algorithm}

Algorithm \ref{alg:greedcover} has a nice approximation ratio as follow.

\bthm\cite{Hochbaum:1996:ACP:241938.241941} Algorithm \ref{alg:greedcover} achieves 
a $1 - \frac{1}{e}$ approxiamtion for optimal solution.
\ethm

However, there are several difficulties which prevent us from utilizing 
Algorithm \ref{alg:greedcover}. (\rmnum{1}): The $1- 1/e$ bound is obtained from the
perspective of cuts, not of edges. Therefore, using this algorithm, we cannot control the number of
edges we need to remove; however the edge budget $l$ is restrict in our problem.
(\rmnum{2}): Even to heuristically employ this algorithm, we have to
enumerate the set $\mathcal{C}$ of length-bounded s-t cuts. We know that the number of cuts is 
exponential to the edge number; then to enumerate the cuts between any 
two vertices is computationally impossible. (\rmnum{3}): Another reasonable angle is to enumerate all
their minimum length-boudned s-t cuts (the number is still exponential \cite{citeulike:9610989}), 
due to their compact representation; however, even to find the minimum length-bounded 
s-t cuts is proven as $\mathcal{A}\mathcal{P}\mathcal{X}$-Complete\cite{flowwithpathrestrictions}.
Given above difficulties, instead of using cut, we will tackle our problem 
from {\it short path} perspective in the rest of this paper.
}



\comment{
3.1 Notations
Expected Number (EN) of infection nodes from source node s notated as N(s).
Truncated distance Expected Number (TdEN) of infection nodes from source node s notated as Nk(s), where k is the truncated distance. The distance from node s to infection node should be no larger than k.

3.2 Problem Definition
	we propose a contrainted edge-cut manner to minimize the average EN and TdEN. 
formulation 1:
		cut(edges)=argminsNN(s)   s.t. number of cut edges L			(1)
Since the interaction network is modeled as uncertain graph, then N(s) can also be written as:
					N(s)=tNp(s,t)				(2)
where p(s,t) indicats the probability of infection from node s to node t. 
Therefore, the dual optimal formulation from eqn.(1) can be rewritten as:
cut(edges)=argminsNtNp(s,t)+*(number of cut edges) 	(3)
formulation 2:
	Similar with formulation 1, the TdEN type of optimal equation is:
			cut(edges)=argminsNtNpk(s,t)+*(number of cut edges)	(4)
	where pk(s,t) indicats the probability of infection from node s to node t within k steps.

	When given the optimal-cut edges, It can be seen that which kinds of relationship removed could 
	efficiently slow down the disease spreading rate. Then, we could leverage the information to make decision.
}

%% file: optimization.tex
\section{Path Algebra and Optimization Algorithm} 
\label{section:optimization} 

In this section, we introduce a numerical relaxation approach to solve the Boolean function maximization problem (and thus the  de-small-world problem). 
Here, the basic idea is that since the direct solution for the Boolean function maximization problem is hard, instead of working on the Boolean (binary) edge variable, we relax to it to be a numerical value. 
However, the challenge is that we need to define the numerical function optimization problem such that it meet the following two criteria: 1) it is rather accurately match the Boolean function maximization; and 2) it can enable numerical solvers to be applied to optimize the numerical function. 
In Subsection~\ref{pathalgebra}, we introduce the numerical optimization problem based on the path algebra. 
In Subsection~\ref{optimization}, we discuss the optimization approach for solving this problem. 
 
\subsection{Path-Algebra and Numerical Optimization Problem}
\label{pathalgebra}

\begin{figure} [!htp] 
\begin{center} 
\centering 
\subfigure{ 
\includegraphics[height=1in,width=1.2 in]{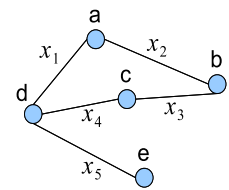}  
} 
\caption{Algebra Variable}  
\label{fig:alg-sym}  
\end{center}   
\end{figure}

To construct a numerical optimization problem for the Boolean function maximization format of the de-small-world problem, we introduce the following path-algebra to describe all the short paths between any reachable pair in $\mathcal{R}_G$. 
For each edge $e$ in the graph $G=(V,E)$, we associate it with a variable $x_e$. 
Then, for any reachable pair $(u,v) \in \mathcal{R}_G$, we define its corresponding path-algebra expression $\mathcal{P}(u,v)$ as follows: 
\beqnarr
\mathcal{P}(u,v)=\sum_{p \in P(u,v)} \prod_{e \in p} x_e
\eeqnarr
Taking the path-algebra for $(b,d)$ in Figure~\ref{fig:runningexampleandor} and Figure~\ref{fig:alg-sym} as example, we have 
\beqnarr
\mathcal{P}(b,d)=x_2 x_1 + x_3 x_4 \nonumber
\eeqnarr 

Basically, the path-algebra expression $\mathcal{P}(u,v)$ directly corresponds to the Boolean expression $I(u,v)$ by replacing $AND (\wedge)$  with product ($\times$), $OR (\vee)$ with sum ($+$), and Boolean variable $e$ with algebraic variable $x_e$.  
Intuitively, $\mathcal{P}(u,v)$ records the weighted sum of each path in $P(u,v)$, where the weight is the product based on the edge variable $x_e$.  
Note that when $x_e=1$ for every edge $e$, when $\mathcal{P}(u,v)$ simply records the number of different short paths (with length no more than $k$) between $u$ and $v$, i.e., $\mathcal{P}(u,v)=|P(u,v)|$. 
Furthermore, if assuming $x_e \geq 0$, then $\mathcal{P}(u,v)=0$ is equivalent to in each path $p \in P(u,v)$, there is at least one edge variable is equivalent to $0$. 
In other words, assuming if variable $x_e=0$ iff $e=T$, then $\mathcal{P}(u,v) =0 $ iff $I(u,v)=F$ and $\mathcal{P}(u,v)>0$ iff $I(u,v)=T$. 

Given this, we may be tempted to optimize the follow objective function based on the path-algebra expression to represent the Boolean function maximization problem: \\$\sum_{(u,v) \in \mathcal{R}_G} \mathcal{P}(u,v)$.
However, this does not accurately reflect our goal, as to minimize $\sum_{(u,v) \in \mathcal{R}_G} \mathcal{P}(u,v)$, we may not need any $\mathcal{P}(u,v)=0$ (which shall be our main goal).
This is because $\mathcal{P}(u,v)$ corresponds to the weighted sum of path products. 
Can we use the path-algebra to address the importance of $\mathcal{P}(u,v)=0$ in the objective function? 

We provide a positive answer to this problem by utilizing an exponential function transformation. 
Specifically, we introduce the following {\em numerical maximization} problem based on the path expression: 
\beqnarr
\sum_{(u,v) \in \mathcal{R}_G} e^{-\lambda \mathcal{P}(u,v)}, \mbox{\  where}, 0 \leq x_e \leq 1, \sum x_e \geq X-L
\label{eqn:goal}
\eeqnarr
Note that $0 \leq e^{-\lambda \mathcal{P}(u,v)} \leq 1$ (each $x_e \geq 0$), and only when $\mathcal{P}(u,v)=0$, $e^{-\lambda \mathcal{P}(u,v)}=1$ (the largest value for each term). 
When $\mathcal{P}(u,v)  \approx 1$, the term $e^{-\lambda \mathcal{P}(u,v)}$ can be rather small (approach $0$). 
The parameter $\lambda$ is the adjusting parameter to help control the exponential curve and smooth the objective function. 
Furthermore, the summation constraint $\sum x_e \geq X-L)$ is to express the budget condition that there shall have $L$ variables with $x_i \approx 0$. 
Here $X$ is the total number of  variables in the objective function ($X=|E|$ if we consider every single edge variable $x_e$). 
\comment{
Finally, we note that here we do not consider the constraint that there shall have $L$ variables with $x_i \approx 0$. 
We will address this problem during our numerical optimization procedure using , where we initially set each $x_e=1$.}


\subsection{Gradient Optimization} 
\label{optimization}

Clearly, it is very hard to find the exact (or closed form) solution for maximizing function in Equation 4 under these  linear constraints. 
In this section, we utilize the standard {\em gradient} (ascent) approach together with the {\em active set} method ~\cite{Hager:2006:NAS} to discover a local maximum. 
The gradient ascent takes steps proportional to the positive of the gradient iteratively to approach a local minimum. 
The active set approach is a standard approach in optimization which deals with the {\em feasible regions} (represented as constraints). Here we utilize it to handle the constraint in Equation 4. 

\noindent{\bf Gradient Computation:} 
To perform gradient ascent optimization, we need compute the gradient $g(x_e)$ for each variable $x_e$. 
Fortunately, we can derive a closed form of $g(x_e)$ in $\sum_{(u,v) \in \mathcal{R}_G} e^{-\lambda \mathcal{P}(u,v)}$ as follows: 
\begin{small} 
\beqnarr 
g(x_e) = \frac{\partial \sum_{(u,v) \in \mathcal{R}_G} e^{-\lambda \mathcal{P}(u,v)}}{\partial x_e} \nonumber  
= \sum_{(u,v) \in \mathcal{R}_G} -\lambda \mathcal{P}(u, v, e) e^{-\lambda \mathcal{P}(u,v)},
\label{eqn:grd} 
\eeqnarr
\end{small}
where $\mathcal{P}(u, v, e)$ is the sum of the path-product on all the paths going through $e$ and we treat $x_e=1$ in the path-product. 
More precisely, let $P(u,v,e)$ be the set of all short paths (with length no more than $k$) between $u$ and $v$ going through edge $e$, and then, 
\beqnarr
\mathcal{P}(u,v,e)=\sum_{p \in P(u,v,e)} \prod_{e^\prime \in p \setminus \{e\}} x_{e^\prime}  
\label{eqPe}
\eeqnarr
Using the example in Figure~\ref{fig:runningexampleandor} and Figure~\ref{fig:alg-sym}, we have  
\beqnarr
\mathcal{P}(b,d,e_1)=x_2 \nonumber
\eeqnarr

Note that once we have all the gradients for each edge variable $x_e$, then we update them accordingly,  
\begin{eqnarray*} 
x_e = x_e +  \beta g(x_e), 
\end{eqnarray*} 
where $\beta$ is the step size (a very small positive real value) to control the rate of convergence. 
 
\noindent{\bf $\mathcal{P}(u,v)$ and $\mathcal{P}(u,v,e)$ Computation}
To compute the gradient, we need compute all  $\mathcal{P}(u,v)$ and $\mathcal{P}(u,v,e)$ for $(u,v) \in \mathcal{R}_G$. 
Especially, the difficulty is that even compute the total number of simple short paths (with length no more than $k$) between $u$ and $v$, denoted as $|P(u,v)|$ is known to be expensive.  
In the following, we describe an efficient procedure to  compute $\mathcal{P}(u,v)$ and $\mathcal{P}(u,v,e)$ efficiently. 
The basic idea is that we perform a DFS from each vertex $u$  with traversal depth no more than $k$. 
During the traversal form vertex $u$, we maintain the partial sum of both $\mathcal{P}(u,v)$ and $\mathcal{P}(u,v,e)$ for each $v$ and $e$ where $u$ can reach within $k$ steps. 
After each traversal, we can then compute the exact value of $\mathcal{P}(u,*)$ and $\mathcal{P}(u,*,*)$. 

\input{Figures/ComputePUVE.tex}

The DFS procedure starting from $u$ to compute all $\mathcal{P}(u,*)$ and $\mathcal{P}(u,*,*)$ is illustrated in Algorithm \ref{alg:computepuv}. 
In Algorithm \ref{alg:computepuv}, we maintain the current path (based on the DFS traversal procedure) in $p$ and its corresponding product $\sum_{e \in p} x_e$ is maintained in variable $w$ (Line $9$ and $10$). 
Then, we incrementally update $\mathcal{P}(u,v)$ assuming $v$ is the end of the path $p$ (Line $11$). 
In addition, we go over each edge in the current path, and incrementally update $\mathcal{P}(u,v)$ ($w/x_e=\prod_{e^\prime \in p \setminus \{e\}} x_{e^\prime}$, Line $13$.) 
Note that we need invoke this procedure for every vertex $u$ to compute all $\mathcal{P}(u,v)$ and $\mathcal{P}(u,v,e)$. 
Thus, the overall time complexity can be written as $O(|V|\overline{d}^k)$ for a random graph where $\overline{d}$ is the average vertex degree. 

\comment{
 the short path numbers for the reachable pairs containing $u$ are computed 
in a DFS prodecure. Applying this procedure to all the vertices in $V$, then we can get all the $P(u, v)$ 
and $P(u, v, e_i)$. Note that sometime to store all the $P(u, v, e_i)$ for all the edges is spatially 
prohibitive(in the worst case $O(mn^2), m=|E|, n=|V|$), in which case we do not record $P(u, v, e_i)$,  
and instead we calculate $P(u, v, e_i)$ dynamically when calculating $g(e_i)$ which only needs one more  
round of DFS. In both cases the time complexity for this step is}

\noindent{\bf Overall Gradient Algorithm}
\input{Figures/OptimizationAlg.tex}

The overall gradient optimization algorithm is depicted in Algorithm \ref{alg:optimization}. 
Here, we use $\mathcal{C}$ to describe all the edges which need be processed for optimization. 
At this point, we consider all the edges and thus $\mathcal{C}=E$. Later, we will consider to first select some candidate edges. 
The entire algorithm performs iteratively and each iteration has three major steps: 

\noindent{\bf Step 1} (Lines $6-8$): it calculates the gradient $g(x_e)$ of for every edge variable $x_e$ and an average gradient $\overline{g}$; 

\noindent{\bf Step 2} (Lines $9-16$): only those variables are not in the active set $\mathcal{A}$ will be updated. 
Specifically, if the condition  ($\sum_{e \in E} x_e \geq |E|-L$) is not met, we try to adjust $x_e$ back to the feasible region. Note that by using $g(x_e)-\overline{g}$ (Line $11$) instead of $g(x_e)$ (Line $13$), we are able to increase the value of those $x_e$ whose gradient is below average. 
However, such adjustment can still guarantee the overall objective function is not decreased (thus will converge). 
Also, we make sure $x_e$ will be between $0$ and $1$. 

\noindent{\bf Step 3} (Lines $17-22$): the active set is updated. When an edge variable reaches $0$ or $1$, we put them in the active set so that we will not need to update them in Step 2. However, for those edges variables in the active set, if their gradients are less (higher) than the average gradient for $x_e=0$ ($x_e=1$), we will release them  from the active set and let them to be further updated. 
 
Note that the gradient ascent with the active set method guarantees  the convergence of the algorithm (mainly because the overall objective function is not decreased). 
However, we note that in Algorithm~\ref{alg:optimization}, the bounded condition ($\sum_{e \in E} x_e \geq |E|-L$) may not be necessarily satisfied even with the update in Line $11$. 
Though this can be achieved through additional adjustment, we do not consider them mainly due to the goal here is not to find the exact optimization, but mainly on identifying the smallest $L$ edges based on $x_e$.  
Finally, the overall time complexity of the optimization algorithm is $O(t(|V|*\overline{d}^k+|E|))$,  
given $t$ is the maximum number of iterations before convergence.

%% file: Figures/ComputePUVE.tex
\begin{algorithm}[ht] 
\caption{{\bf ComputePUVE}$(G, u, k, p, w)$}  
\label{alg:computepuv} 
\begin{algorithmic}[1] 
\STATE {\bf Input}: $G=(V, E)$ and starting vertex $u$; 
\STATE {\bf Input}: spreading parameter $k$, path $p$, and partial product $w$; 
\STATE {\bf Output}: $P(u, *)$, $P(u, *, *)$; 
\IF { $|p| = k$ \COMMENT{traversal depth no more than $k$} }
    \RETURN
\ENDIF 
\STATE $z = s.top()$ \COMMENT{the last visited vertex in the traveral}
\FORALL {$v \in Neighbor(z)$ and $v \notin p$ \COMMENT{simple path}} 
    \STATE $p.push(v)$ \COMMENT{the current path}; 
    \STATE $w \leftarrow w \times x_{(v,z)}$ \COMMENT{corresponding path product}; 
    \STATE $\mathcal{P}(u,v) \leftarrow \mathcal{P}(u,v)+w$; 
    \FORALL {$e \in p$ \COMMENT{every edge in the current path}} 
       \STATE $\mathcal{P}(u,v,e) \leftarrow \mathcal{P}(u,v,e) + \frac{w}{x_e}$; 
    \ENDFOR
   \STATE {\bf ComputePUVE}$(G, u, k, p, w)$; 
   \STATE $p.pop()$; $w \leftarrow \frac{w}{x_{(v,z)}}$; 
\ENDFOR 
\end{algorithmic} 
\end{algorithm}

%% file: Figures/OptimizationAlg.tex
\begin{algorithm}[ht] 
\caption{${\bf OptimizationAlg}(G, L)$}  
\label{alg:optimization} 
\begin{algorithmic}[1] 
\STATE {\bf Input}: $G=(V, E)$, and edge removal budget $L$; 
\STATE {\bf Output}: edge set $E_r$; 
\STATE $\forall e \in \mathcal{C}$ ($\mathcal{C}=E$), $x_e \leftarrow 1$; \COMMENT{initialization}
\STATE $\mathcal{A} \leftarrow \emptyset$; \COMMENT{active set}
\WHILE{NOT every $x_e$ converges} 
\STATE $\forall x_e$, calculate $\mathcal{P}(u, v)$ and $\mathcal{P}(u, v, e)$ using Algorithm \ref{alg:computepuv};
\STATE $\forall x_e$, $g(x_e) \leftarrow -\lambda \sum_{T(x_e)} e^{-\lambda \mathcal{P}(u,v)}\mathcal{P}(u,v,e)$; 
\STATE $\overline{g} \leftarrow \frac{\sum_{x_e \in \mathcal{C} \setminus \mathcal{A}} g(x_e)}{| \mathcal{C} \setminus \mathcal{A}|}$ \COMMENT{average gradient}; 
\FORALL { $e \in \mathcal{C} \setminus \mathcal{A}$}
    \IF { bound reached ($\sum_{e \in \mathcal{C}} x_e < |\mathcal{C}|-L$) \COMMENT{using $x_e$ from last iteration}}
        \STATE $x_e \leftarrow max(x_e+\beta(g(x_e)-\overline{g}), 0)$;
    \ELSE
        \STATE $x_e \leftarrow max(x_e+\beta g(x_e), 0)$;
    \ENDIF
    \STATE $x_e \leftarrow min(x_e, 1)$;
\ENDFOR
\FORALL { $e \in \mathcal{C} \setminus \mathcal{A}$ and ($x_e = 1$ or $x_e = 0$)}
     \STATE $\mathcal{A} \leftarrow \mathcal{A} \cup \{e\}$; \COMMENT{add to active set}
\ENDFOR
\FORALL{ $e \in \mathcal{A}$ and ($(x_e=0 \wedge g(x_e)<= \overline{g}) \vee (x_e=1 \wedge g(e)>= \overline{g})$)}
    \STATE $\mathcal{A} \leftarrow \mathcal{A} \backslash \{e\}$; \COMMENT{remove from active set}
\ENDFOR
\ENDWHILE 
\STATE sort all $\{x_e\}$ in increasing order, and add top $L$ edges to $E_r$; 
\end{algorithmic} 
\end{algorithm}

%% file: speedup.tex
\section{Short Betweenness and Speedup Techniques} 
\label{section:scalable} 
In Section \ref{section:optimization}, we reformulate our problem into a numerical optimization problem. We further develop an iterative {\it gradient} algorithm to select the top $L$ edges in to $E_r$. However, the basic algorithm can not scale well to very large graphs due to the large number ($|E|$) of variables  involved. 
In this section, we introduce a new variant of the edge-betweenness and use it to quickly reduce the variables needed in the optimization algorithm (Algorithm~\ref{alg:optimization}). 
In addition, we can further speedup the DFS procedure to compute $\mathcal{P}(u,v)$ and $\mathcal{P}(u,v,e)$ in Algorithm~\ref{alg:computepuv}. 

\subsection{Short Betweenness} 
\label{shortbetweenness}

In this subsection, we consider the following question: {\em What edge importance measure can directly correlate with $x_e$ in the objective function in Eq. 4 so that we can use it to help quickly identify a candidate edge set for the numerical optimization described in Algorithm~\ref{alg:optimization}}?
In this work, we propose a new edge-betweenness measure, referred to as the {\em short betweenness} to address the this question. 
It is intuitively simple and has an interestingly relationship with respect to the gradient $g(x_e)$ for each edge variable.
It can even be directly applied for selecting $E_r$ using the generic procedure in Section~\ref{problem} and  is much more effective compared with the global and local edge-betweenness which measure  the edge importance in terms of the shortest path (See comparison in Section~\ref{section:exp}). 

Here we formally define $\mathcal{r}(e_i)$ as {\it short betweenness}. 
\bdefin{\bf (Short Betweenness:)} The short betweenness $SB(e)$ for edge $e$ is as follows, 
$SB(e) = \sum_{(u, v) \in \mathcal{R}_G} \frac{|P(u,v,e)|}{|P(u,v)|}$. 
\edefin 
Recall that $(u,v) \in \mathcal{R}_G$ means $d(u,v) \leq k$;  $|P(u,v)|$ is the number of short paths between $u$ and $v$; and $|P(u,v,e)|$ is the number of short paths between $u$ and $v$ which must go through edge $e$. 
The following lemma highlights the relationship between the short betweenness and the  gradient of edge variable $x_e$:

\blemma
Assuming for all edge variables $x_e=1$, then $g(x_e) \geq -SB(e)$. 
\elemma 
\bproof
\beqnarr
g(e) & = & \sum_{(u,v) \in \mathcal{G}_R} -\lambda \mathcal{P}(u, v, e) e^ {-\lambda \mathcal{P}(u,v)}  \nonumber \\
       & = & \sum_{ (u,v) \in \mathcal{G}_R} -\lambda  |P(u, v, e)| e^{-\lambda |P(u,v)|}  (\forall e, x_e=1) \nonumber \\
& \geq&\sum_{ (u,v) \in \mathcal{G}_R} \frac{-\lambda |P(u, v, e)|}{\lambda |P(u,v)|} (e^{-x} < 1/x, x>0) \nonumber \\
& =&  -SB(e) \nonumber
\eeqnarr
\eproof

Basically, when $x_e=1$ for every edge variable $x_e$ (this is also the initialization of Algorithm~\ref{alg:optimization}), the (negative) short betweenness serves a lower bound of the gradient $g(e)$. Especially, since the gradient is  negative, the higher the gradient of $|g(e)|$ is, the more likely it can maximize the objective function (cut more reachable pairs in $\mathcal{R}_G$. 
Here, the short betweenness $SB(e)$ thus provide an upper bound (or approximation) on $|g(e)|$ (assuming all other edges are presented in the graph); and measures the the edge potential in removing those local reachable pairs. 
Finally, we note that Algorithm~\ref{alg:computepuv} can be utilized to compute 
$|P(u,v)|$ and $|P(u,v,e)|$, and thus the short betweenness (just assuming $x_e=1$ for all edge variables).

\noindent{\bf Scaling Optimization using Short Betweenness:}
First, we can directly utilize the short betweenness to help us pickup a candidate set of edge variables, and then Algorithm~\ref{alg:optimization} only need to work on these edge variables (considering other edge variables are set as $1$). 
Basically, we can choose a subset of edges $E_s$ which has the highest short betweenness in the entire graph. The size of $E_s$ has to be larger than $L$; in general, we can assume $|E_s|=\alpha L$, where $\alpha>1$. In the experimental evaluation (Section~\ref{section:exp}), we found when $\alpha=5$, the performance of using candidate set is almost as good as the original algorithm which uses the entire edge variables. 
Once the candidate set edge set is selected, we make the following simple observation: 

\blemma
Given a candidate edge set $E_s \subseteq E$, if any reachable pair $(u,v) \in \mathcal{R}_G$ can be cut by $E_r$ where $E_r \subseteq E_s$ and $|E_r|=L$, then, each path in $P(u,v)$ must contains at least one edge in $E_s$.  
\elemma

Clearly, if there is one path in $P(u,v)$ does not contain an edge in $E_s$, it will always linked no matter how we select $E_r$ and thus cannot cut by $E_r \subseteq E_s$.
In other words, $(u,v)$ has to be cut by $E_s$ if it can be cut by $E_r$. 
Given this, we introduce $\mathcal{R}_s=\mathcal{R}_G \subseteq \mathcal{R}_{G \setminus E_s}$. 
Note that $\mathcal{R}_s$ can be easily computed by the DFS traversal procedure similar to Algorithm~\ref{alg:computepuv}.
Thus, we can focus on optimizing 
\beqnarr
\sum_{(u,v) \in \mathcal{R}_s} e^{-\lambda \mathcal{P}(u,v)}, \mbox{\  where}, 0 \leq x_e \leq 1, \sum x_e \geq X-L
\label{eqn:goal2}
\eeqnarr
Furthermore, let $E_P=\bigcup_{(u,v) \in  \mathcal{R}_s} \bigcup_{p \in P(u,v)} p$, which records those edges appearing in certain path linking a reachable pair cut by $E_P$. 
Clearly, for those edges in $E \setminus E_P$, we can simply prune them from the original graph $G$ without affecting the final results.  
To sum, the short betweenness measure can help speed up the numerical optimization process by  reducing the number of edge variables and pruning non-essential edges from the original graph.

%% file: exp.tex
\vspace*{-3.0ex}
\section{Experimental Study}
\label{section:exp}
In this section, we report the results of the empirical study of our methods. 
Specifically, we are interested in the performance (in terms of reachable pair cut) and the efficiency (running time). 

\begin{figure*}[tbp]
\vspace{-3ex}
\resizebox{\textwidth}{!} {
    \begin{minipage}[c]{0.3\textwidth}
	\caption{Network Statistics}
	{\small
	\begin{tabular}{|l|r|r|r|}
	\hline
	Dataset & \multicolumn{1}{c|}{$|V|$} & \multicolumn{1}{c|}{$|E|$} & \multicolumn{1}{c|}{$\pi$}  \\ \hline

	Gnutella04 & 10,876 & 39,994 & 9 \\ \hline
	Gnutella05 & 8,846  & 31,839 & 9 \\ \hline
	Gnutella06 & 8,717 & 31,525 & 9 \\ \hline
	Gnutella08 & 6,301 & 20,777 & 9 \\ \hline
	Gnutella09 & 8,114  & 26,013 & 9 \\ \hline
	Gnutella24 & 26,518 & 65,369 & 10 \\ \hline
	Gnutella25 & 22,687 & 54,705 & 11 \\ \hline
	Gnutella30 & 36,682  & 88,328 & 10 \\ \hline
	Gnutella31 & 62,586 & 147,892 & 11 \\ \hline
	\end{tabular}
	}
	\label{fig:network_info}
    \end{minipage}
    \hspace{0.02cm}

    \begin{minipage}[c]{0.7\textwidth}
		\begin{center}
		\centering
		\includegraphics[height=1.8in,width=5in]{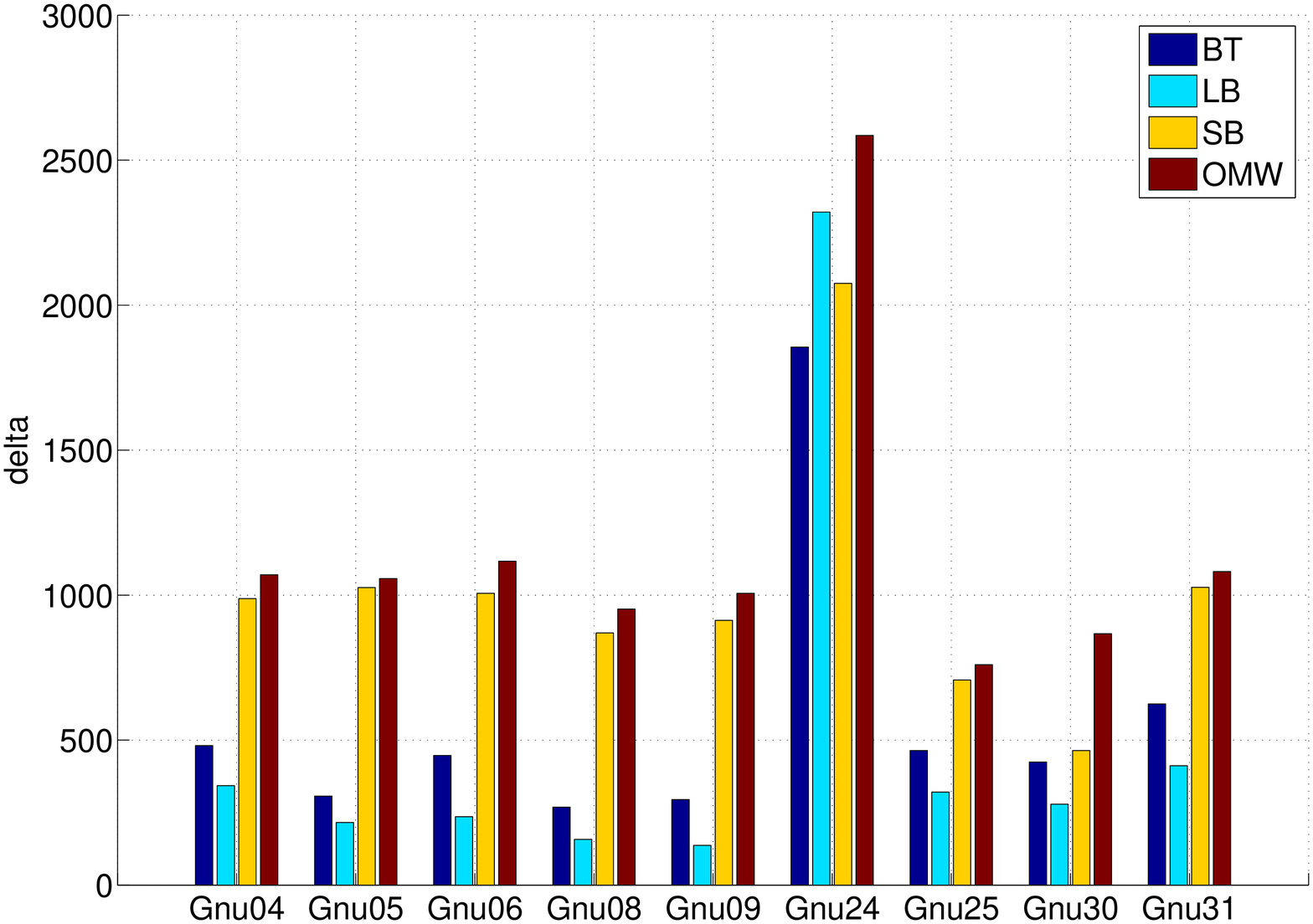} 
		\label{fig:histogram} 
		\caption{$\delta$ for all real datasets} 
		\end{center}	
	\end{minipage}
    \hspace{0.02cm}	
}
\vspace{-3.0ex}
\end{figure*}

\noindent{\bf Performance:} 
Given a set of edges $E_r$ with budget $L$, the total number of reachable pairs being cut by $E_r$ is
$|\mathcal{R}_G|-|\mathcal{R}_{G \setminus E_r}|$ or simply $\Delta|\mathcal{R}_G|$.
We use the average pair being cut by an edge, i.e., $\delta=\frac{\Delta|\mathcal{R}_G|}{L}$ as the performance measure. 

\noindent{\bf Efficiency:} The running time of different algorithms.

\comment{With respect to $\delta$, there are three important parameters. 1) $|V|$: 
given the graph with certain structures, that is, small-world graphs (small graph diameter and large clustering 
coefficient), the first factor effecting the algorithm performance is graph size $|V|$. How does
a graph with a different vertex size respond when its edges are removed?  2) $L$: the edge budget 
controls the number of edges that be removed by each algorithm. Of course, larger $L$ is, more reachable
pairs could be removed. Thus, when can we get the best $\delta$ with respect to different $L$? 
3) $k$: the maximum length of short paths determines the number of total reachable pairs. Clearly,
larger $k$ is, larger the number of reachable pairs in $G$. Here we want to know whether 
$\delta$ is sensitive to $k$ or not?
}

\noindent{\bf Methods:} Here we compare the following methods: 

1) {\it Betweenness based method}, which is defined in terms of the shortest paths between 
any two vertices in the whole graph $G$; hereafter, we use $BT$ to denote the method based 
on this criterion. 

2) {\it Local Betweenness based method }, which, compared with betweenness method($BT$), takes only the
vertex pair within certain distance into consideration; hereafter, we use $LB$ to stand for
the method based on local betweenness.

3) {\it Short Betweenness based method}, the new betweenness introduced in this paper, which considers all short paths whose length is no more than certain threshold.  Here we denote the method based on short betweenness as $SB$. 

4) {\it Numerical Optimization method}, which solves the de-small-world problem iteratively by calculating
gradients and updating the edge variables $x_e$. Based on whether the method use the candidate set or not,
we have two versions of optimization methods: $OMW$ (Optimization Method With candidate set) and 
$OMO$ (Optimization Method withOut candidate set). Note that we normally choose the top $5L$ edges as our
candidate set.

\begin{figure} [!htp]
\begin{center}
\centering
\includegraphics[height=1.5 in,width=3in]{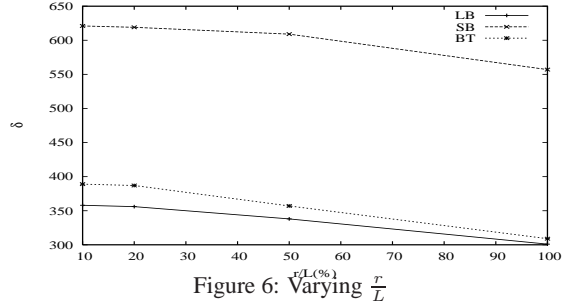} 
\label{fig:r} 
\vspace*{-3.0ex}
\caption{Varying $\frac{r}{L}$} 
\end{center}  
\vspace*{-4.0ex}
\end{figure}

As mentioned before in Section \ref{problem}, we have a generic procedure to select $L$ edges depending
on parameter $r$ (batch size). We found for different methods $BT$, $LB$ and $SB$, the effects of $r$ seem to be rather  small (as illustrated in Figure \ref{fig:r}). Thus, in the reminder of the experiments, we choose $r=L$, i.e., we select the top $L$ edges using the betweenness calculated for the entire (original graph). 

All the algorithms are implemented using C++ and the Standard Template Library (STL), and the experiments are conducted on a 2.0GHz Dual Core
AMD Opteron CPU with 4.0GB RAM running on Linux.

\subsection{Result on Synthetic Datasets }

In this subsection, we study the performance and efficiency of different methods on the synthetic datasets. 
Here, we generate various synthetic networks from two well-known small-world models: 
{\it Watts and Strogatz model} (WS model) \cite{citeulike:99} 
generating small-world graphs by interpolating between ER graph and a regular ring lattice;
the small-world model proposed by Kleinberg \cite{citeulike:3158434} (KS model). 
Then, the networks generated from WS model, KS model are referred to as the WS network and KS network, respectively. 

\begin{figure*} [!htp]
\begin{center}
\centering
\mbox{
\subfigure[Varying $|V|$ for KS]{
\includegraphics[height=1.2 in,width=1.5in]{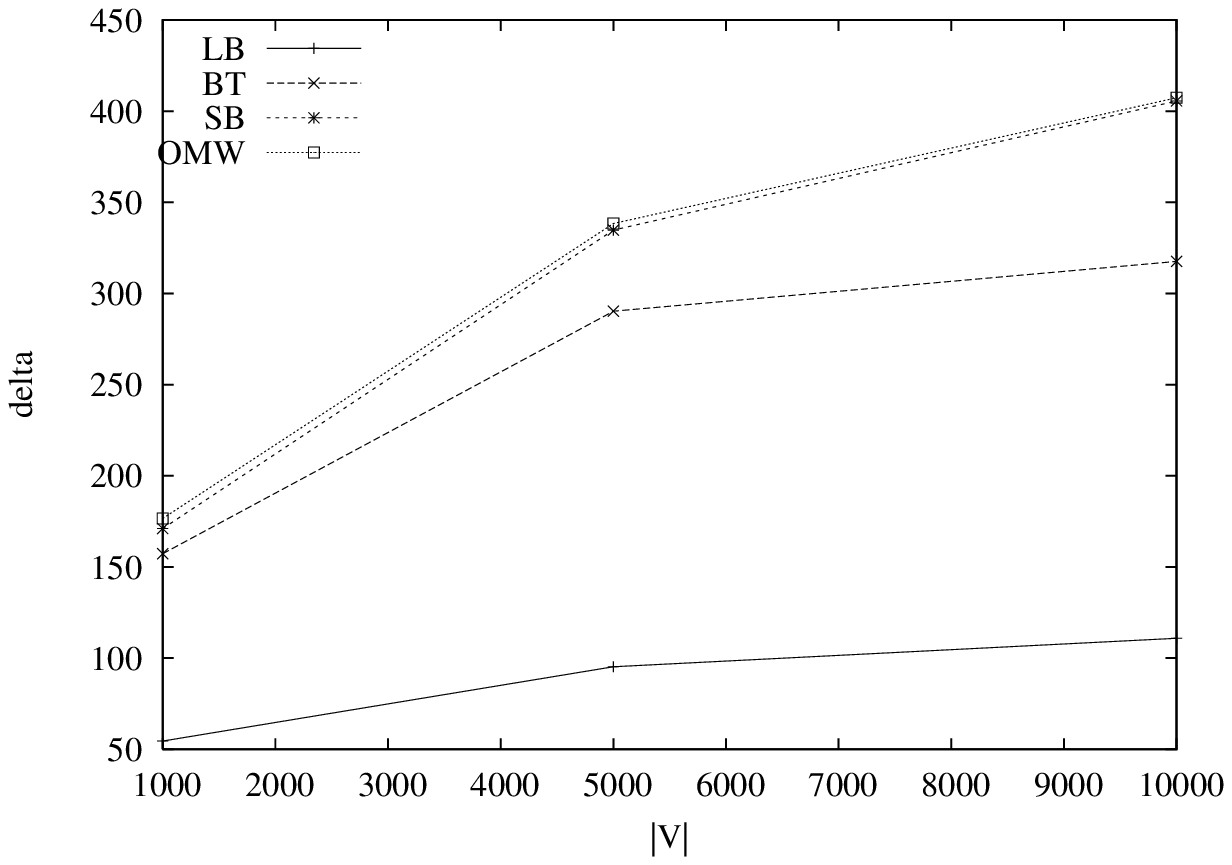} 
\label{fig:varying_size_ks} 
}
\subfigure[Varying $L$ for KS]{  
\includegraphics[height=1.2 in,width=1.5in]{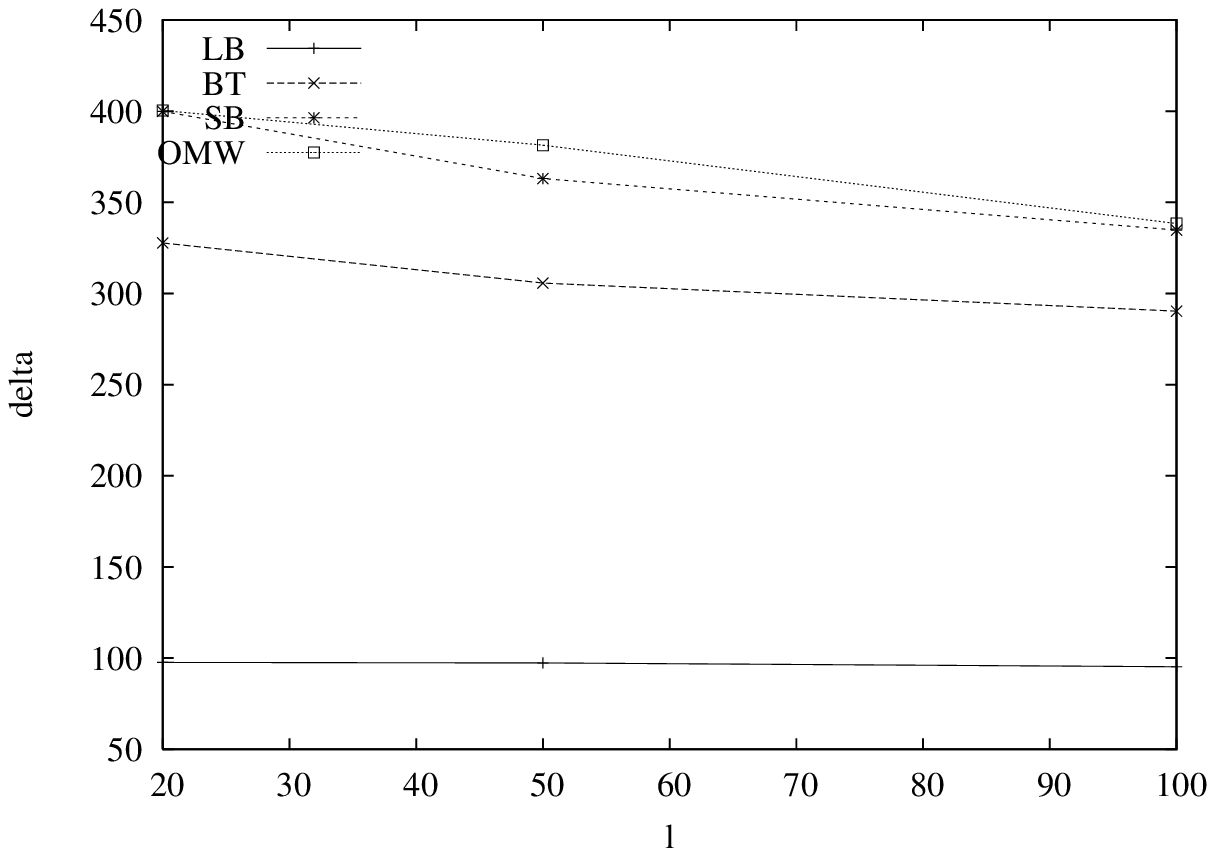}  
\label{fig:varying_l_ks} 
}
\subfigure[Varying $k$ for KS]{
\includegraphics[height=1.2 in,width=1.5in]{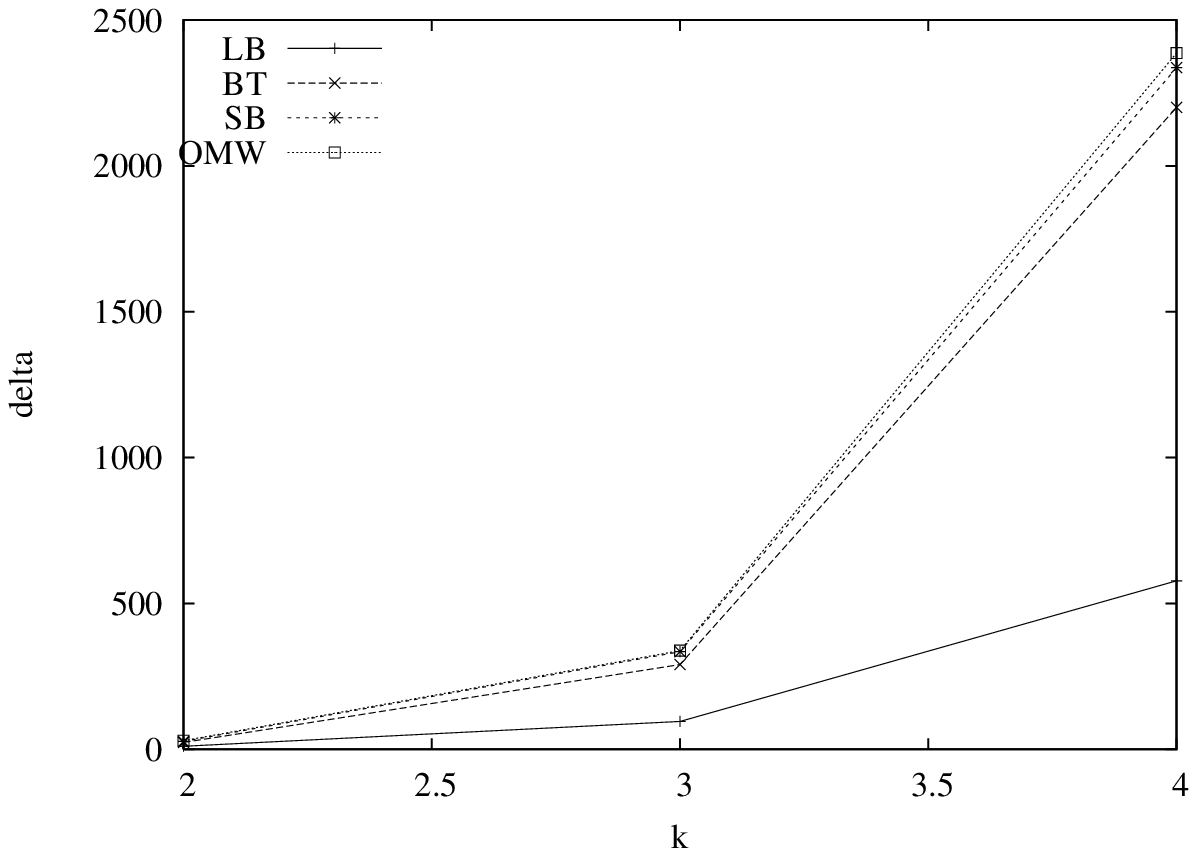} 
\label{fig:varying_k_ks} 
}
\subfigure[Running Time for KS]{
\includegraphics[height=1.2 in,width=1.5in]{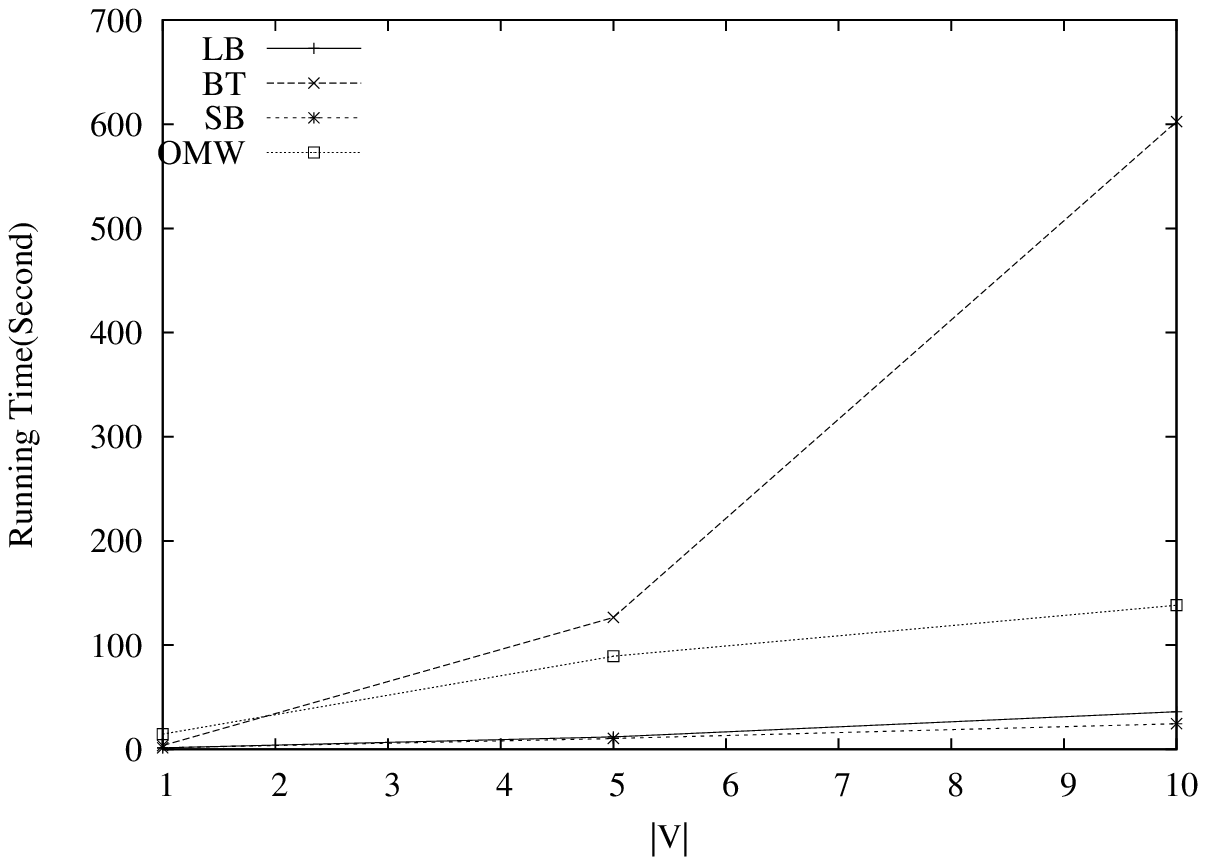} 
\label{fig:time_ks} 
}
}
\mbox{
\subfigure[Varying $|V|$ for WS]{  
\includegraphics[height=1.2in,width=1.5in]{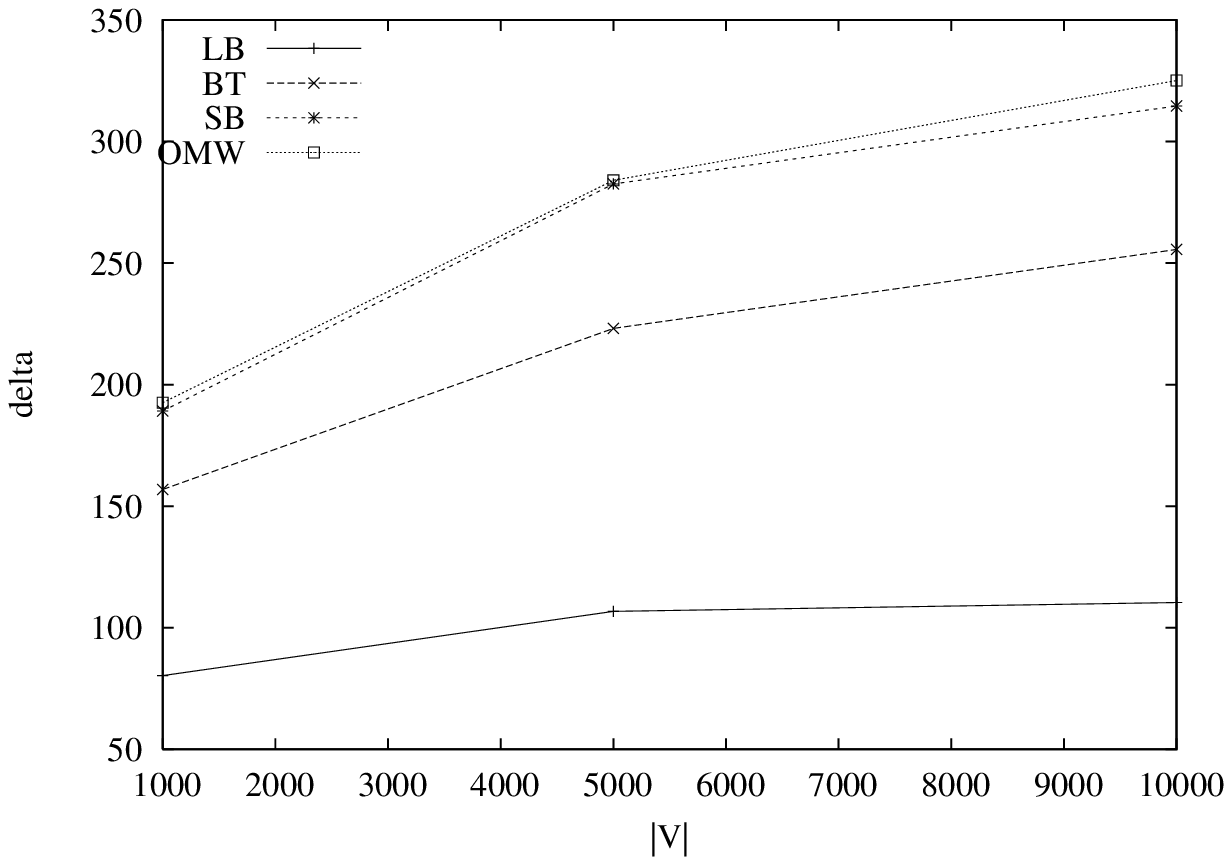}  
\label{fig:varying_size_ws} 
}
\subfigure[Varying $L$ for WS]{
\includegraphics[height=1.2 in,width=1.5in]{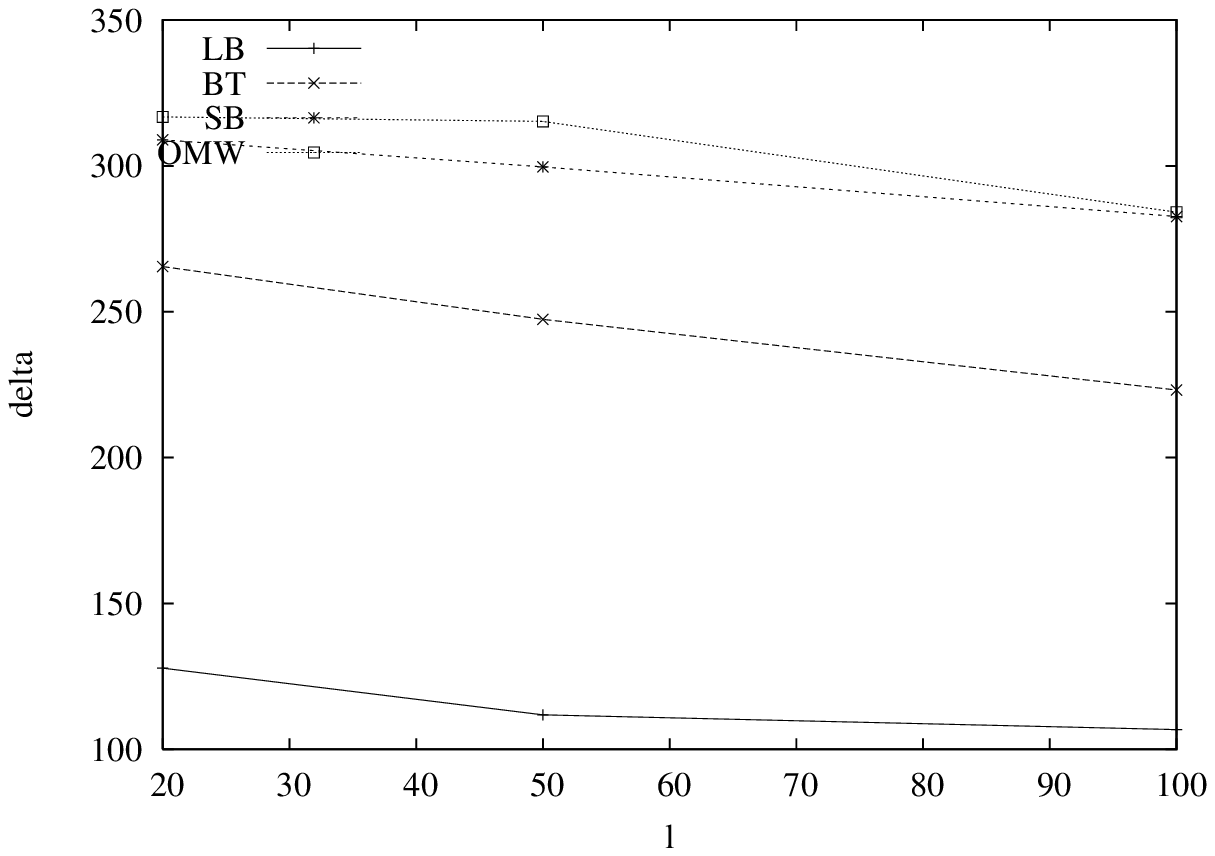} 
\label{fig:varying_l_ws} 
}
\subfigure[Varying $k$ for WS]{  
\includegraphics[height=1.2in,width=1.5in]{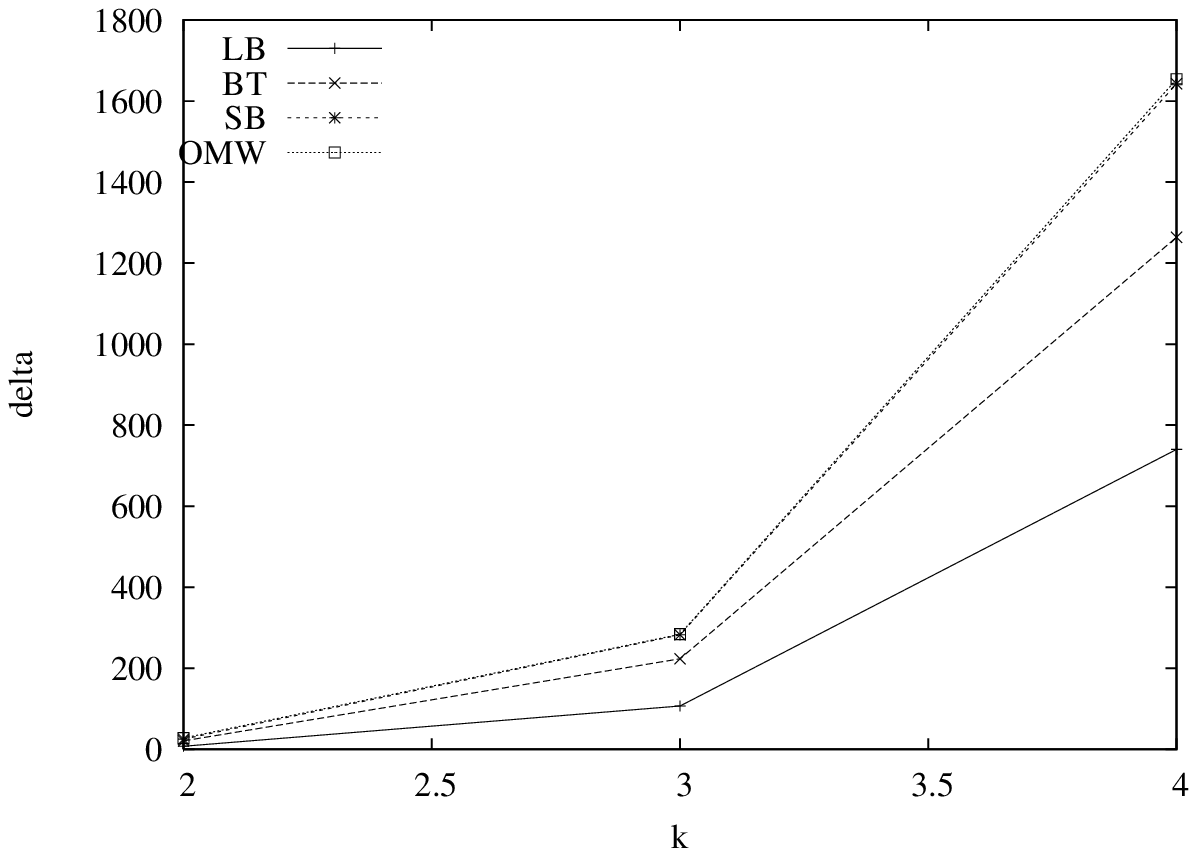}  
\label{fig:varying_k_ws} 
}
\subfigure[Running Time for KS]{
\includegraphics[height=1.2 in,width=1.5in]{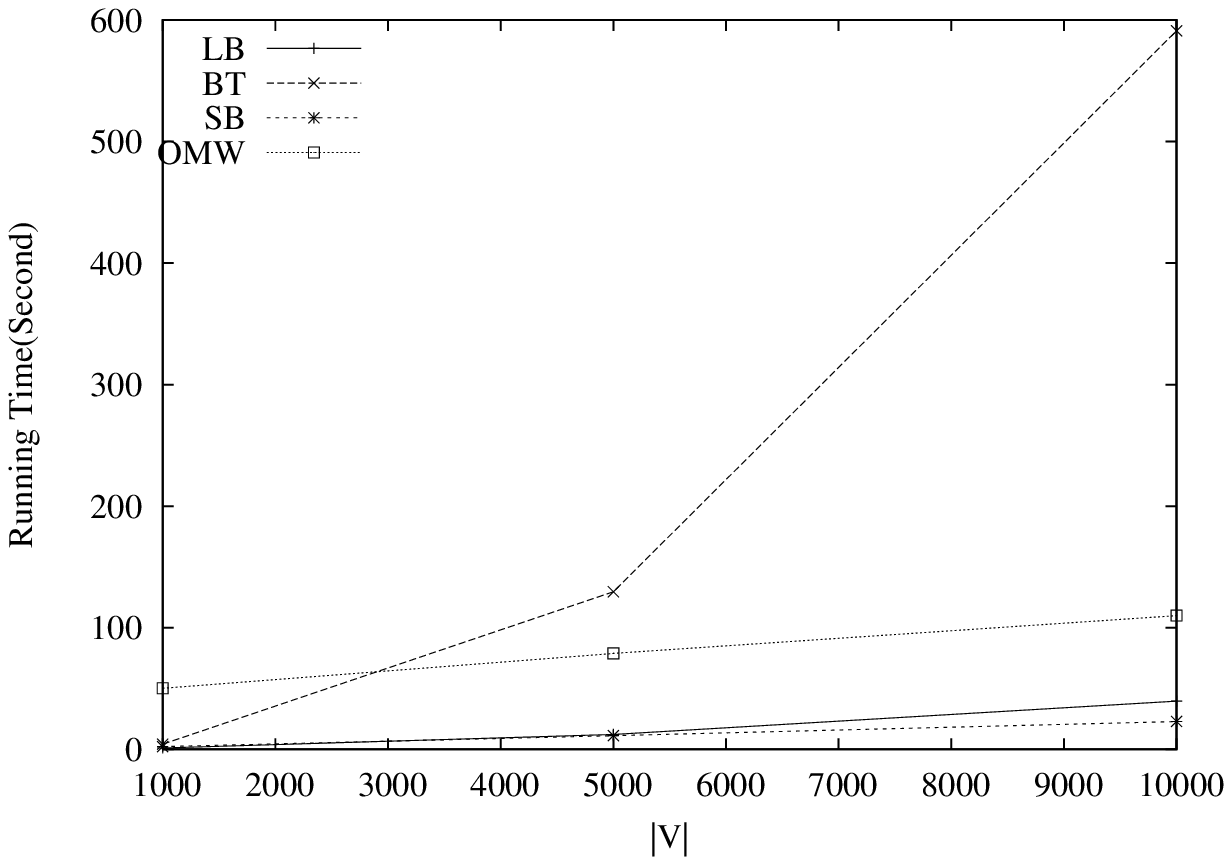} 
\label{fig:time_ws} 
}
}
\caption{Experiments on Synthetic Datasets} 
\end{center}  
\end{figure*}

In the following, we conducted three groups of experiments.
\noindent{\bf Varying $|V|$:} In this group of experiments, we generate networks respectively with 
the two models (WS, KS) using vertex size $1k$, $5k$ and $10k$. 
We also set the edge budget $L = 1000$ (edge removal budget)  and $k=3$ (spreading parameter).
The results are summarized in 
Figure \ref{fig:varying_size_ks} and \ref{fig:varying_size_ws}.
From these two figures, we can see that $LB$ method always 
produces the worst result (its $\delta$ is around $100$, meaning each edge on average contribute to around $100$ reachable pairs). 
Meantime, $\delta$ for $BT$ method increases dramatically from $150$ to $300$ for both $KS$ and $WS$
graphs. Comparatively, $OMW$ always reduces the biggest number of pairs compared with other methods. 
More specifically, its $\delta$ grows from $175$ to more than $400$. Meanwhile, $SB$ method produces
the similar result as $OMW$ method.
This suggests the power of short betweenness (which directly forms an upper bound for the absolute gradient $g(x_e)$). 

\noindent{\bf Varying $L$:} In this group of experiments, we study the reduction effect for different
$L$ and the result for $KS$ model is reported in Figure \ref{fig:varying_l_ks}. Generally, with the 
increase of $L$, $\delta$ decreases. This is reasonable because more reachable pairs is
removed, each edge can remove the smaller number of reachable pairs. For the specific algorithms, 
similar to the situation in last group of experiment, $LB$ and $OMW$ methods produce 
the lowest $\delta$ and highest $\delta$, respectively.
Then the number of reduced reachable pairs by $BT$ method is about three times that of $LB$, and 
is about $\frac{3}{4}$ of that reduced by $SB$ and $OMW$. These cases also happen for the graphs generated
by $WS$ model as in Figure \ref{fig:varying_l_ws}.

\noindent{\bf Varying $k$:} Remember that we define the short path as the paths with length at most $k$. 
Given $G$, obviously $k$ determines the size of reachable pairs. Given different $k$, the result of all 
algorithms are reported in Figure \ref{fig:varying_k_ks} for $KS$ model graphs. We can see that
generally, with the increase of $k$, the strength of each edge($\delta$) increases. This is understandable
because with $k$ increasing, each edge could effect more reachable pairs. For the specific algorithms,
$LB$ produces the lowest $\delta$ for all $k$. Then other three methods produce similar $\delta$, which 
are normally about four times between than $LB$. The similar situation happens for $WS$ graphs as 
in Figure \ref{fig:varying_k_ws}.

\subsection{Result on Real Datasets } 

In this subsection, we study the performance of our algorithms on real datasets. 
The benchmarking datasets are listed in Figure~\ref{fig:network_info}. 
All networks contain certain properties commonly observed in social networks, such as small diameter.
All datasets are downloadable from Stanford Large Network Dataset Collection~\footnote{http://snap.stanford.edu/data/index.html}.

In Figure~\ref{fig:network_info}, we present important characteristics of all real datasets,
where $\pi$ is graph diamter. All these nine networks are snapshots of the Gnutella peer to
peer file sharing network starting from August 2002. Nodes stand for the hosts in the Gnutella
network topology and the edges for the connections between the hosts.



\input{Figures/Table.tex}

\noindent{\bf Varying $L$:} We perform this group of experiments on dataset $Gnu05$ and we 
fix $k =3$. Here we run these methods on three different edge buget $L$: $500$, $1000$ and $2000$.
The result is reported in Table \ref{tab:varying_l}. The general trend is that with smaller
$L$, $\delta$ becomes bigger. This is because the set of reachable pairs removed by different edges
could have intersection; when one edge is removed, the set of reachable pairs for other edges is 
also reduced. For particular methods, $BT$ and $OMO$ methods produces the lowest and highest
$\delta$, and the different between $OMW$ and $OMO$ is very small.

\noindent{\bf Varying $k$:} In this group of experiments, we fix $L = 1000$ and we choose 
$Gnu04$. Here we choose three values for $k$: $2$, $3$ and $4$. 
The result is reported in Table \ref{tab:varying_k}. From the result, we can see that
when $k$ becomes bigger, $\delta$ become higher. This is also reasonable:when
$k$ becomes bigger, more reachable pairs are generated and meanwhile $|E|$ is constant;
therefore, each edge is potentially able to remove more reachable pairs. From the above three
groups of experiments, we can see that $OMO$ does not produce significant results compared with
$OMW$. Therefore, in the following experiment, we do not study $OMO$ method again.

\noindent{\bf $\delta $ on all real datasets:} In this groups of experiment, we study the performance
of each method on these nine datasets, with $L$ being proportional to $|E|$. Specifically, $L = |E|\times 1\%$.
We report the result in Figure \ref{fig:histogram}. $LB$ generally produces the lowest $\delta$, around 
half that of $BT$; and also the best method, is the $SB$ and $OMW$ methods.
Specifically, $OMW$ is always slightly better than $SB$.


\comment{

\noindent {\bf Facebook:} a social network crawled from Facebook. Each node represents a user and there is a directed edge from node $i$ to 
node $j$ if user $i$ posted message on user $j$'s personal page.

\noindent {\bf Slashdot:} a technology-related social network. It contains user-to-user friendship links which was collected in November 2008.

\noindent {\bf BerkStan:} a web graph describing the hyperlinks from pages of berkely.edu to pages of stanford.edu.
This data was collected in 2002 and presented in~\cite{DBLP:journals/corr/abs-0810-1355}.

part of Google Programming Contest.

\noindent {\bf Youtube:} a YouTube video-sharing network containing around $1.1$ million users and $5$ millions user-to-user links.

\noindent {\bf As-skitter:} a internet topology graph from traceroutes run daily in 2005.

\noindent {\bf Flickr:} a Flickr photo-sharing network with around $1.8$ million users and $22$ millions links, which is crawled in 2007.

\noindent {\bf Flickr-growth:} another Flickr social networks used to investigate the behavior of social network's growth 
in~\cite{mislove-2008-flickr}.

\noindent {\bf WikiTalk:} a Wikipedia users' communication network. This network contains all the users and discussion from the inception of 
Wikipedia till January 2008.
In the network, each node represents one user and there is a directed edge from node $i$ to node $j$ if user $i$ edited a talk page of 
user $j$ for communication.

\noindent {\bf Orkut:} an online social network owned by Google. Each node denotes one user and each undirected link identifies friendship 
between two users.


\noindent {\bf LiveJournal:} a fraction of LiveJournal social network. Each node is a highly active user and undirected edges are used to 
identify the friendship between different users.

a undirected edge between node $i$ and node $j$ if certain relationship between them is identified by EntityCube.
}

%% file: Figures/Table.tex
\begin{figure*}[tbp]
\vspace{-3ex}
\resizebox{\textwidth}{!} {
    \begin{minipage}[l]{0.31\textwidth}
		\caption{Running Time (Seconds)}
		{\scriptsize
		\begin{tabular}{|c|c|c|c|c|}
		\hline
		Time & $BT$ & $LB$ & $SB$ & OMW \\ 
		\hline		
		10,876&	382.27&	24.82&	33.75&	1021.66\\
		\hline	
		8,846&	21346.54&	496.17&	8.98&	110.80\\
		\hline
		62586&	392.54&	25.31&	34.60&	1092.55\\
		\hline
		\end{tabular}
		}
	\label{tab:run_time}
    \end{minipage}
    \hspace{0.01cm}
	
    \begin{minipage}[c]{0.3\textwidth}
		\caption{$\delta$ By Varying $l$}
		\label{tab:varying_l}
		{\scriptsize
		\begin{tabular}{|c|c|c|c|c|c|}
		\hline
		$l$ & $BT$ & $LB$ & $SB$ & OMW & OMO \\ 
		\hline		
		500&	240&	415&	912&	996& 973\\ \hline
		1000&	261&	372&	740&	803& 805\\ \hline
		2000&	301&	329&	572&	620& 622\\ \hline
		\end{tabular}
		}
    \end{minipage}

    \begin{minipage}[r]{0.3\textwidth}
		\caption{$\delta$ By Varying $k$}
		\label{tab:varying_k}
		{\scriptsize
		\begin{tabular}{|c|c|c|c|c|c|}
		\hline
		$k$ & $BT$ & $LB$ & $SB$ & OMW & OMO \\ 
		\hline		
		2&	25&		32&		55&	58& 58\\ \hline
		3&	261&	372&	740&	803& 805\\ \hline
		4&	761&	976&	2113&	2389 & -\\ \hline
		\end{tabular}
		}
    \end{minipage}
}
\vspace{-3.0ex}
\end{figure*}

%% file: conclude.tex
\section{Conclusion}
\label{section:con}

In this paper, we introduce the {\it de-small-world} network problem; to solve it, we first present 
a greedy edge betweenness based approach as the benchmark and then provide
a numerical relaxation approach to slove our problem using OR-AND boolean format, which
can find a local minimum. In addition, we introduce the {\it short-betweenness} to speed up our algorithm. 
The empirical study demonstrates the efficiency and effectiveness of our approaches. In the future, we plan
to utilize MapReduce framework(e.g. Hadoop) to scale our methods to handle graphs with 
tens of millions of vertices. 


%% file: Appendix.tex
\appendix

\section{Proof of Theorem 1}

To prove this theorem, we first introduce the {\it dense $\kappa$-sub-hypergraph} problem.
Let $H=(V_H,E_H)$ be a hypergraph, where $V_H$ is the vertex set, and $E_H$ is the hyperedge set, such that $e_h \in E_h$ and $e_h \subseteq V$. Each hyperedge in the hypergraph is a subset of vertices (not necessarily a pair as in the general graph). 
Furthermore, given a subset of vertices $V_s \subseteq V_H$, if an hypergraph $e_h \subseteq V_S$, then, we say this hyperedge is {\em covered} by the vertex subset $V_s$. 

\bdefin{\bf Dense $\kappa$-Sub-Hypergraph Problem:}
Given hypergraph $G=(V_H, E_H)$ and a parameter $\kappa$, we seek to find a subset of vertices $V_s \subseteq V_H$ and $|V_s|=\kappa$, such that the maximal number of hyperedges in $E_H$ is covered by $V_s$. 
\edefin
Dense $\kappa$-hyper-subgraph problem can be easily proven to be NP-Hard, because its special case, dense $\kappa$-subgraph problem (each edge $e \subset V \times V$) has been shown to be NP-Hard~\cite{Feige99thedense}.  

\begin{figure} [!htp]
\begin{center}
\centering
\includegraphics[height=1 in,width=2.5in]{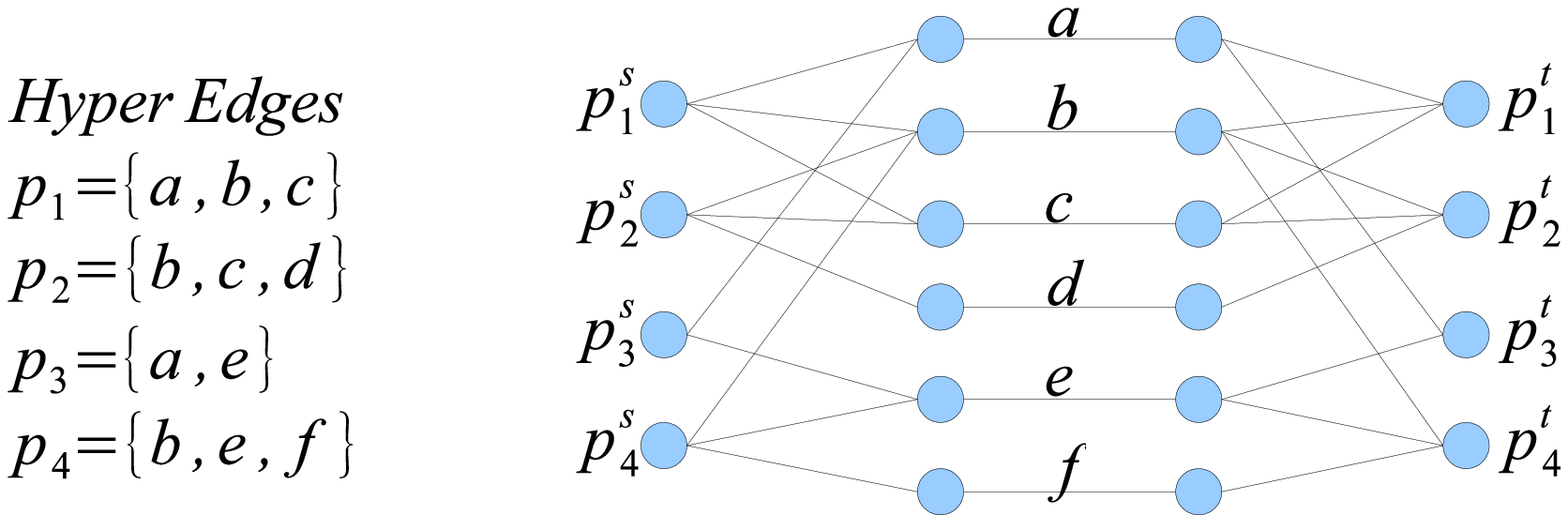} 
\caption{Graph for Reachable Pair cut Problem}
\label{fig:hardness} 
\end{center}  
\end{figure}

\bproof
To reduce  the Dense $\kappa$-Sub-Hypergraph problem to our problem, we construct the following graph from the given hypergraph $H=(V_H,E_H)$. Figure\ref{fig:hardness} illustrates the transformation. 
For each hyperedge $e_i \in E_H$, which consists a set $p_i=\{v_{i_1}, v_{i_2}, \cdots, v_{i_k}\}$ of vertices in $V_H$, we represent it as a vertex pair $p_i^s$ and $p_i^t$  in the graph $G$, and each pair is connected by $i_k$ different paths with length $3$, where each middle edge in $G$ corresponds to a unique vertex in $V_H$. 
For instance for hyperedge $p_1=\{a,b,c\}$, the middle edges of the three-paths in $G$ correspond to edges $a$, $b$, $c$.
To facilitate our discussion, the middle edges of these length-3 paths linking the vertex pairs corresponding to each hypergraph in $H$ are referred to $MS$, which has one-to-one correspondence to the vertex set $V_H$ in the hypergraph. 

Now we show for the dense $\kappa$ sub-hypergraph problem, its optimal solution can be solved by an instance of the general reachable pair cut problem, where $L=\kappa$ and $RS$ consists of all the ($p_i^s,p_i^e$) reachable pairs ($k=3$). In other worlds, $|RS|=|E_H|$. 
Specifically, we need show that if a subset of edges $S$ ($|S|=L$) in $MS$ can maximally disconnect the reachable pairs in $RS$, then its corresponding vertex subset $V_S$ can maximally cover the hyperedges. 
This is easy to observe due to the one-to-one correspondence relationship between $MS$ and $V_H$ and between $RS$ and $E_H$. 
Given this, we need show that the optimal solution of the  reachable pair cut problem can be always found using only edges in $MS$.
Suppose the edge set $ES^\prime$ with size $L$  is the optimal solution which contain some edge $e \in ES$, and $e \notin MS$. 
In this case, we can simply replace $e$ with its adjacent middle edge $e^\prime$ from $MS$ in the result set. 
This is because the replacement $ES^\prime \backslash \{e\} \cup \{e^\prime \}$ will still be able to disconnect all the reachable pairs in $RS$ being cut by $ES^\prime$.
Note that the middle edge $e^\prime$ can cut more paths than $e$ (form a superset of the paths cut by $e$) with respect to $RS$. 
\eproof